\documentclass[11pt,aps,notitlepage,nofootinbib]{revtex4-1}
\usepackage{amssymb,amsmath}
\usepackage{bm}
\usepackage[pdftex]{graphicx}
\usepackage{subfig}

\usepackage[dvipsnames]{xcolor}

\newcommand{\beq}{\begin{equation}}
\newcommand{\beqn}{\begin{eqnarray}}
\newcommand{\eeq}{\end{equation}}
\newcommand{\eeqn}{\end{eqnarray}}
\newcommand{\calO}{{\cal{O}}}
\newcommand{\calP}{{\cal{P}}}
\newcommand{\nc}{\newcommand}
\nc{\ba}{\begin{eqnarray}}
\nc{\ea}{\end{eqnarray}}

\begin{document}

\title{Relativistic Corrections to Nonrelativistic Effective
Field Theories}
\author{Mohammad Hossein Namjoo, Alan H. Guth, and David I. Kaiser}
\email{namjoo@mit.edu ; guth@ctp.mit.edu ; dikaiser@mit.edu }
\affiliation{Department of Physics, 
Massachusetts Institute of Technology, Cambridge, Massachusetts
02139 USA}
\date{\today}
\begin{abstract} 
In this paper we develop a formalism for studying the
nonrelativistic limit of relativistic field theories in a
systematic way. By introducing a simple, nonlocal field
redefinition, we transform a given relativistic theory,
describing a real, self-interacting scalar field, into an
equivalent theory, describing a complex scalar field that encodes
at each time both the original field and its conjugate momentum.
Our low-energy effective theory incorporates relativistic
corrections to the kinetic energy as well as the backreaction of
fast-oscillating terms on the behavior of the dominant, slowly
varying component of the field. Possible applications of our new
approach include axion dark matter, though the methods developed
here should be applicable to the low-energy limits of other field
theories as well. \\

Preprint MIT-CTP/4955
\end{abstract}

\maketitle

\section{Introduction} 

Understanding the nature of dark matter remains a major challenge
at the intersections of astrophysics, cosmology, and particle
physics. Multiple lines of observational evidence indicate that
dark matter should be plentiful throughout the universe,
contributing roughly five times more to the energy density of the
universe than ordinary (baryonic) matter, and that the dark
matter should be cold and collisionless \cite{DMreviews}. From
the standpoint of particle theory, the puzzle of dark matter
includes at least two components: identifying a plausible
dark-matter candidate within realistic models of particle
physics, and developing an accurate, theoretical description that
is suitable for low-energy phenomena associated with cold matter.
For the latter goal, it is important to develop a means of
characterizing the nonrelativistic limit of relativistic quantum
field theories in a systematic way.

In this paper we develop an effective field theory approach for
the nonrelativistic limit of relativistic field theories. We
focus on the nonrelativistic behavior of scalar fields in
Minkowski spacetime, incorporating systematic relativistic
corrections as well as corrections from the fields' (weakly
coupled) self-interactions. We have in mind applications to axion
dark matter \cite{axionDM1,axionDM2,axionDM3,axionDM4}, though
the methods developed here should be applicable to the low-energy
limits of other field theories, such as QED, QCD
\cite{Stewart1,Stewart2,Brambilla05}, and condensed-matter
systems, and to specific phenomena such as ``oscillons''
\cite{GleiserOscillons,AminOscillons, MasakiOscillons} and
``superradiance" \cite{Superradiance1,Superradiance2}. 

Axions are an attractive candidate for dark matter. The
hypothetical particles were originally introduced to address the
strong CP problem in QCD \cite{axionCP1,axionCP2,axionCP3}, but
their expected mass and self-coupling make them well-suited for
cold dark matter as well. In particular, axions, with \hbox{$m
\sim 10^{-4}$--$10^{-5}$} eV, are expected to be produced early
in cosmic history, around the time of the QCD phase transition. 
At that time the typical wavenumber was $k \sim H_{\rm QCD} \sim
10^{-11}$ eV \cite{GHPW15}, where $H_{\rm QCD} \sim T_{\rm QCD}^2
/ M_{\rm pl}$ is the Hubble scale at the time of the QCD
transition, $T_{\rm QCD} \sim 0.1$ GeV is the temperature at
which the transition occurs, and $M_{\rm pl}
\equiv 1/\sqrt{8\pi G} \sim 10^{18}$ GeV is the reduced Planck
mass. Axions are expected to become accelerated gravitationally
during large-scale structure formation, up to speeds $v / c \sim
{\cal O} (10^{-3} )$ within galactic halos \cite{GHPW15}. Hence
even at late times they should remain squarely within the
nonrelativistic limit, albeit in a regime in which relativistic
corrections may be competitive with other nontrivial corrections,
such as from higher-order interaction terms. 

In the nonrelativistic limit, fluctuations that oscillate rapidly
on time scales $m^{-1}$ (where $m$ is the mass of the scalar
field) may be expected to average to zero over time scales
$\Delta t \gg m^{-1}$. However, nonlinear self-couplings can
induce a backreaction of the fast-oscillating terms on the
dominant, slowly varying component of the nonrelativistic field,
affecting its behavior. Such coupling of fast- and
slow-oscillating terms produces measurable effects in many
physical systems, such as neuronal processes related to memory
formation \cite{memory1,memory2}. We develop an iterative,
perturbative procedure to incorporate this backreaction.

Our approach complements other recent work, such as
Refs.~\cite{MasakiOscillons}, \cite{GHPW15}, and \cite{Braaten16}. 
In particular, the authors of
Ref.~\cite{Braaten16} develop a nonrelativistic effective field
theory for describing axions by calculating scattering amplitudes
for $n \rightarrow n$ body scattering in the full, relativistic
theory, and then taking the low-energy limit of those amplitudes
to match coefficients in a series expansion of the effective
potential. In Ref.~\cite{MasakiOscillons}, the authors
developed an effective field theory by using diagrammatic
techniques to integrate out the high-momentum modes. In this
paper, we develop an effective description for the
nonrelativistic limit with no need to calculate scattering
amplitudes in the corresponding relativistic theory. Our method
yields the same coefficients for the leading higher-order
interaction terms in the low-energy limit as those found in
Refs.~\cite{MasakiOscillons} and \cite{Braaten16}, while also
incorporating systematic relativistic corrections which do not
appear in the previous analysis.

In Section \ref{sec:fieldredef} we introduce a convenient field
redefinition to relate the real-valued scalar field described by
the relativistic theory to a complex scalar field more
appropriate to the nonrelativistic limit. In particular, we
introduce a nonlocal field redefinition rather than the local
redefinition that one typically finds in the literature. Although
we expect the resulting descriptions of the low-energy limit to
be equivalent in either redefinition, we find that the nonlocal
redefinition considerably simplifies the derivation. In Section
\ref{sec:nonrelEFT} we develop an effective field theory for the
nonrelativistic field, which incorporates contributions from
fast-oscillating terms on the evolution of the principal, slowly
varying portion of the field. Concluding remarks follow in
Section \ref{sec:conclusions}. In Appendix \ref{sec:canonical} we
demonstrate that our transformation from the real-valued
relativistic field $\phi (t, {\bf x})$ to the complex-valued
nonrelativistic field $\psi (t, {\bf x})$ can be constructed as a
canonical transformation. In Appendix \ref{sec:local_redef} we
demonstrate that the low-energy effective field theory based on
our nonlocal field redefinition explicitly matches what one would
calculate with the local field redefinition in two regimes of
interest. In Appendix \ref{sec:Masaki_compare} we compare our
results with the recent calculation in
Ref.~\cite{MasakiOscillons} and demonstrate that the two results
are equivalent, related to each other by a field redefinition.

\section{Field Redefinition For a Nonrelativistic Formulation}
\label{sec:fieldredef}

In this section we begin with the Lagrangian for a relativistic
field theory that describes a scalar field in Minkowski
spacetime, and introduce a suitable field redefinition with which
we may consider the nonrelativistic limit systematically. Our
goal is to obtain an expression for the Lagrangian that yields
the Schr\"{o}dinger equation as the effective equation of motion
for the redefined field in the extreme nonrelativistic limit. 

We consider a relativistic scalar field of mass $m$ with a
$\lambda \phi^4$ interaction, described by the Lagrangian density
\beq
{\cal L} =- \frac{1}{2} \eta^{\mu\nu} \partial_\mu \phi
\partial_\nu \phi - \frac{1}{2} m^2 \phi^2 - \frac{1}{4!} \lambda
\phi^4 ,
\label{Lrel}
\eeq
where $\phi$ is a real-valued scalar field, and we take the
Minkowski metric to be $\eta_{\mu\nu} = {\rm diag} [-1, 1, 1,
1]$.  The canonical momentum field is $\pi = \dot \phi$, where an
overdot denotes a derivative with respect to time, and the
Hamiltonian is given by
\beq
H = \int d^3 x \, {\cal H}({\bf x}) \ , \qquad {\cal H} =
\frac{1}{2} \pi^2 + \frac{1}{2} ({\bf \nabla} \phi)^2 +
\frac{1}{2} m^2 \phi^2 + \frac{1}{4!} \lambda \phi^4 .
\label{Hrel}
\eeq
The equations of motion take the form
\begin{equation}
\begin{split}
\dot \phi &= {\delta H \over \delta \pi} = \pi ,\\
\dot \pi &= - {\delta  H \over \partial \phi}  = (\nabla^2 - m^2) \phi - \frac{1}{3!} \lambda \phi^3 .
\end{split}
\label{EqRel}
\end{equation}

In contrast, the standard Lagrangian for a free, nonrelativistic
field may be written
\beq
{\cal L} = \frac{i}{2} \left( \dot{\psi} \psi^* - \psi
\dot{\psi}^* \right) - \frac{1}{2m} \nabla \psi \nabla \psi^* ,
\label{Lnonrel}
\eeq
where $\psi$ is a complex scalar field, and overdots again denote
derivatives with respect to time. The kinetic term is normalized
so that the field $\psi$ obeys the standard Poisson bracket
relations with $\psi^*$, so that when quantized the two fields
obey the standard commutation relations, as given below in
Eq.~(\ref{commutator}) and discussed in
Eqs.~(\ref{kinetic})--(\ref{commutator2}). Whereas the Lagrangian
in Eq.~(\ref{Lrel}) yields a second-order equation of motion for
$\phi$, the Lagrangian in Eq.~(\ref{Lnonrel}) yields first order
equations of motion for the real and imaginary parts of $\psi$.
The Lagrangian in Eq.~(\ref{Lnonrel}) has a global $U(1)$
symmetry; the associated conserved charge is simply the number of
particles,
\beq
N = \int d^3 x \> \vert \psi \vert^2 ,
\label{N}
\eeq
confirming the usual expectation that particle number should be
conserved in a nonrelativistic theory, appropriate for energy
scales $E \ll m$. 

Previous authors (see, e.g., Refs.~\cite{GHPW15} and
\cite{Braaten16}) have typically related the nonrelativistic
field $\psi$ to the relativistic field $\phi$ by using the
relations
\begin{subequations}
\begin{align}
\phi (t, {\bf x} )&= \frac{1}{\sqrt{2m} } \left[ e^{-imt} \psi (t, {\bf x}) + e^{imt} \psi^*  (t, {\bf x} ) \right] , \\
\pi (t, {\bf x} ) &= - i \sqrt{ \frac{ m}{2} } \, \left[ e^{-imt} \psi (t, {\bf x}) - e^{imt} \psi^*  (t, {\bf x} ) \right] . 
\end{align}
\label{phitrad}
\end{subequations}
Note that the quantities in brackets in Eqs.~(\ref{phitrad}a) and
(\ref{phitrad}b) could have been written as ${\rm Re} [ e^{-imt}
\psi ]$ and ${\rm Im} [e^{-imt} \psi]$, respectively, so the
equations are independent. Eq.~(\ref{phitrad}b) is not ordinarily
written explicitly, but this equation or something similar has to
be assumed if $\psi(t, {\bf x})$ is to be uniquely defined. At
any fixed time $t$, Eqs.~(\ref{phitrad}a) and (\ref{phitrad}b),
{\it taken together}, give a one-to-one mapping between the
complex-valued $\psi(t, {\bf x})$ and the two real-valued
functions $\phi(t, {\bf x})$ and $\pi(t,{\bf x})$.  If we use
Eqs.~(\ref{EqRel}) and (\ref{phitrad}) to derive equations of
motion for $\psi$, and then set $\lambda=0$ and neglect all
rapidly oscillating terms (proportional to $e^{\pm imt}$), we
find equations of motion consistent with the Lagrangian of
Eq.~(\ref{Lnonrel}). However, for the purpose of systematically
obtaining the relativistic corrections to the nonrelativistic
theory, we find it more convenient to start from a nonlocal field
redefinition. In place of Eqs.~(\ref{phitrad}), we write
\begin{subequations}
\begin{align}
\phi (t, {\bf x}) &= \frac{1}{ \sqrt{ 2m} } {\cal P}^{-1/2} \left[ e^{-imt} \psi (t, {\bf x} ) + e^{imt} \psi^* (t, {\bf x} ) \right] , \\
\pi (t, {\bf x} ) &= - i \sqrt{ \frac{ m}{2} } \, {\cal P}^{1/2} \left[ e^{-imt} \psi (t, {\bf x} ) - e^{imt} \psi^* (t, {\bf x} ) \right] ,
\end{align}
\label{phipsi1}
\end{subequations}
where we have defined
\beq
{\cal P } \equiv \sqrt{ 1 - \frac{ \nabla^2}{m^2} } .
\label{Pdef1}
\eeq
Note that $m {\cal P}$ corresponds to the total energy of a free,
relativistic particle.  Eqs.~(\ref{phipsi1}) can be inverted to
obtain an equation for $\psi$ in terms of $\phi$ and $\pi$:
\beq
\psi(t,{\bf x}) = \sqrt{\frac{m}{2}} e^{i m t} {\cal P}^{1/2} \left(\phi(t,{\bf x}) + {i \over m} {\cal P}^{-1} \pi(t,{\bf x}) \right) .
\label{psieq}
\eeq

Although our field redefinition in Eqs.~(\ref{phipsi1}) involves
nonlocal operators, the new fields $\psi$ and $\psi^*$ are well
behaved in the nonrelativistic limit, in which the operator
${\cal P}$ can be expanded in powers of $\nabla^2 / m^2$.  The
leading term matches the local definitions of
Eq.~(\ref{phitrad}).  Furthermore, even though the $\psi$ field
has a nonlocal relation to $\phi$, the $\psi$ field is local with
respect to itself.  When quantized, the commutator of
$\psi(t,{\bf x})$ and $\psi^*(t,{\bf y})$ becomes
\beqn
\begin{split}
[ \psi(t,{\bf x}), \psi^*(t,{\bf y})] &= \frac{m}{2} {\cal
P}_x^{1/2} {\cal P}_y^{1/2} \left[\phi(t,{\bf x}) - {i \over m}
{\cal P}_x^{-1} \pi(t,{\bf x}) \, , \, \phi(t,{\bf y}) + {i \over
m} {\cal P}_y^{-1} \pi(t,{\bf y})\right] \\ &= \delta^3 ({\bf x}
- {\bf y}) ,
\end{split}
\label{commutator}
\eeqn
where the subscripts on ${\cal P}$ indicate the coordinates on
which they act, and we assumed of course that $[ \phi(t,{\bf x})
\, , \, \pi(t,{\bf y}) ] = i \delta^3 ({\bf x} - {\bf y})$.  The
simplicity of this result is a consequence of the fact that the
nonlocal operators on the two lines of Eq.~(\ref{phipsi1}) are
the inverses of each other.

Given Eqs.~(\ref{EqRel}) and (\ref{psieq}), it is straightforward
to work out the equation of motion for $ \psi(t,{\bf x})$:
\beq
i \dot{\psi} = m \left( {\cal P} - 1 \right) \psi + \frac{
\lambda e^{imt} }{4! \, m^2 } \, {\cal P}^{-1/2} \left[ e^{-imt}
\, {\cal P}^{-1/2} \psi + e^{imt} \, {\cal P}^{-1/2} \psi^*
\right]^3 .
\label{psieom1a}
\eeq
A key step in this calculation, which motivates the nonlocal
operator ${\cal P}$, is the calculation of the time derivative of
$\phi + \frac{i}{m} {\cal P}^{-1} \pi$:
\beqn
\begin{split}
\dot \phi + \frac{i}{m} {\cal P}^{-1} \dot \pi &= \pi + \frac{i}{m} {\cal P}^{-1} \left[ (\nabla^2-m^2) \phi - \frac{1}{3!}\lambda \phi^3 \right]\\
&= \pi - i m {\cal P} \phi - \frac{i \lambda}{3! m} {\cal P}^{-1}
\phi^3 \\ &= - i m {\cal P} \left[\phi + \frac{i}{m}{\cal P}^{-1}
\pi \right] - \frac{i \lambda}{3! m} {\cal P}^{-1} \phi^3 .
\end{split}
\eeqn
The definition of ${\cal P}$ was chosen so that the first term on
the right-hand side of the last line above is proportional to
$\phi + \frac{i}{m} {\cal P}^{-1} \pi \propto \psi $.  If we used
the local definition instead, the right-hand side would have also
contained a term proportional to $e^{2 i m t} \psi^*$, which
would lead to rapidly oscillating terms even in the free field
theory ($\lambda=0$).  With the nonlocal field definition of
Eqs.~(\ref{phipsi1}), the free theory leads to no rapidly
oscillating terms.

It is for some purposes useful to write a Lagrangian density for
the nonrelativistic formulation, so we note that the equation of
motion in Eq.~(\ref{psieom1a}) can be derived from the Lagrangian
density
\beq
{\cal L} = \frac{ i}{2} \left( \dot{\psi} \psi^* - \psi
\dot{\psi}^* \right) - m \psi^* \left( {\cal P} - 1 \right) \psi
- \frac{\lambda}{4 \cdot 4! m^2} \left[ e^{-i m t} {\cal
P}^{-1/2} \psi + e^{i m t} {\cal P}^{-1/2} \psi^* \right]^4 .
\label{Lrelpsi}
\eeq
Note that the free field theory terms in Eq.~(\ref{Lrelpsi}),
corresponding to $\lambda=0$, show a manifest global $U(1)$
symmetry, $\psi \to e^{i \theta} \psi$.  This symmetry is
associated with the conservation of particle number, which is
exact in the free theory even when energies are relativistic.

To construct the corresponding Hamiltonian density, we explicitly
decompose the field into its real and imaginary parts, $\psi
\equiv \psi_R + i \psi_I$.  The kinetic terms of the Lagrangian
density then become
\beq
{\cal L}_{\rm kinetic} = \frac{ i}{2} \left( \dot{\psi} \psi^* -
\psi \dot{\psi}^* \right) = \dot {\psi}_R \psi_I - \psi_R \dot
{\psi}_I .
\label{kinetic}
\eeq
If we take $\psi_R$ to be the canonical field, then $\psi_I$ will
become the canonical momentum $\partial {\cal L}/\partial
\dot{\psi}_R$.  If we proceeded directly to construct the
Hamiltonian density in the standard way, ${\cal H} = \pi
\dot{\psi}_R - {\cal L}$, the second term on the right of
Eq.~\eqref{kinetic} would be rewritten as $- \psi_R \dot{\psi}_I
= - \psi_R \dot \pi$.  This, however, would take us outside the
standard Hamiltonian procedure, in which the Hamiltonian is
assumed to be a function of the fields and their canonical
momenta, but not the time derivatives of canonical momenta.  To
avoid this problem, we can add a total time derivative to the
Lagrangian density, which does not change the equations of
motion.  So we replace ${\cal L}_{\rm kinetic}$ by
\beq
{\cal L}'_{\rm kinetic} = {\cal L}_{\rm kinetic} + \frac{d}{d
t}\left( \psi_R \psi_I \right) = 2 \dot {\psi}_R \psi_I .
\eeq 
To absorb the factor of 2 we find it convenient to define the
canonical field $\psi_c$ to be
\beq
\psi_c(t,{\bf x}) \equiv \sqrt{2} \hskip 1pt \psi_R(t,{\bf x}) ,
\eeq
from which it follows that
\beq
\pi_c(t,{\bf x}) \equiv \frac{\partial {\cal L}}{\partial \dot{\psi}_c(t,{\bf x})} = \sqrt{2} \hskip 1pt \psi_I(t,{\bf x}) .
\eeq
Thus, we have
\beq
\psi(t,{\bf x}) = \frac{1}{\sqrt{2}} \Bigl( \psi_c(t,{\bf x}) + i \pi_c(t,{\bf x}) \Bigr) .
\label{psi_c-pi_c}
\eeq
The Hamiltonian density is then given by
\beq
{\cal H} = \pi_c \dot {\psi}_c - {\cal L} = m \psi^* \left( {\cal
P} - 1 \right) \psi + \frac{\lambda}{4 \cdot 4! \, m^2 } \left[
e^{-imt} \, {\cal P}^{-1/2} \psi + e^{imt} \, {\cal P}^{-1/2}
\psi^* \right]^4 ,
\label{Hamiltonian}
\eeq
where $\psi$ is given by Eq.~\eqref{psi_c-pi_c}. The canonical
quantization of this Hamiltonian would give
\beq
[ \psi_c(t,{\bf x}), \pi_c(t,{\bf y}) ] = i \delta^3 ({\bf x} -
{\bf y}) ,
\label{commutator2}
\eeq
which is equivalent to Eq.~\eqref{commutator}. The discussion
here has shown that the Hamiltonian of Eq.~\eqref{Hamiltonian}
gives the correct equation of motion for the field $\psi(t,{\bf
x})$.  In Appendix~\ref{sec:canonical}, we show that the
transformation from $\phi(t,{\bf x})$ to $\psi(t,{\bf x})$ can be
constructed as a canonical transformation, which guarantees that
the canonical commutators will be preserved, as we have found.

Our findings are in contrast with the recent claim in Ref.
\cite{Banerjee:2018pvs}, in which the authors found it
problematic to derive a Lagrangian for a complex, nonrelativistic
field from a real-valued relativistic one. Based on the canonical
transformation we have explicitly formulated, we do not believe
there should be any difficulty in constructing ${\cal L} (\psi,
\dot{\psi})$. It is interesting to note that we may obtain the
Lagrangian in Eq.~(\ref{Lrelpsi}) directly from the Lagrangian in
Eq.~\eqref{Lrel} by the following procedure. Replace $\dot \phi$
in the Lagrangian by an auxiliary field using $\chi = \dot \phi$,
and impose the relation via a Lagrange multiplier: $\cal{L}[\phi,
\dot \phi] \to \cal{L}[\phi, \chi]+\zeta (\chi-\dot \phi)$, where
$\zeta$ is a Lagrange multiplier. We next note that $\chi$ is a
nondynamical variable in the resulting Lagrangian. We remove it
from the theory by varying ${\cal L}$ with respect to $\chi$,
which yields an expression for $\chi$, which we then substitute
back into ${\cal L}$. We use similar relations as in Eqs.~\eqref{phipsi1} 
to relate the set $\{\phi, \zeta \}$ (instead of
$\{\phi, \pi \}$) to the set $\{ \psi, \psi^* \}$, and substitute
the relations back into the Lagrangian. 

We close this section with a final observation about the field
redefinitions in Eqs.~(\ref{phipsi1}), compared to the
traditional relation in Eqs.~(\ref{phitrad}). The usual relation
of Eqs.~(\ref{phitrad}) has the advantage of being local, but the
resulting Lagrangian --- even for free fields --- would no longer
preserve the manifest $U (1)$ symmetry, thereby obscuring the
fact that such models should conserve particle number. Moreover,
if we had used Eqs.~(\ref{phitrad}), the fast oscillatory factors
$(e^{\pm i mt})$, which are absent in the free field theory using
our nonlocal formulation, would have appeared in the free-field
Lagrangian. We build upon these advantages of the relations in
Eqs.~(\ref{phipsi1}) in the following section.

\section{Effective Field Theory in the Nonrelativistic Limit}
\label{sec:nonrelEFT}

Given the Lagrangian and equation of motion for $\psi$ in
Eqs.~(\ref{Lrelpsi}) and (\ref{psieom1a}), we may now consider
the effective description for such a model in the nonrelativistic
limit. We aim to take the nonrelativistic limit in a way that
enables us to incorporate relativistic corrections
systematically. One step will obviously be to expand the nonlocal
operator ${\cal P}$ in powers of $\nabla^2 /m^2$. However, we
must take additional steps in order to recover an appropriate
description in the nonrelativistic limit. In particular, we must
find some way to incorporate the effects of fast-oscillating
terms on the behavior of the slowly varying field.

Consider, as a first step, the equation of motion of
Eq.~(\ref{psieom1a}) in the limit in which we may ignore all
higher spatial-derivative terms arising from the expansion of
${\cal P}$. Then we find
\beq
i \dot{\psi} \simeq - \frac{1}{2m} \nabla^2 \psi +
\frac{\lambda}{8 m^2} \vert \psi \vert^2 \psi + \frac{
\lambda}{4! \, m^2} \left[ e^{-2imt} \psi^3 + e^{4imt} \psi^{* 3}
+ 3 e^{2imt} \vert \psi \vert^2 \psi^* \right] .
\label{psieom1}
\eeq
The first two terms on the right-hand side consist of the usual
terms in the Schr\"{o}dinger equation for a model incorporating
self-interactions \cite{GHPW15,Braaten16}. The remaining terms
are usually neglected in the nonrelativistic limit, because they
are proportional to fast-oscillating factors, and in the limit of
large $m$, such terms might be expected to average to zero if the
system were observed over time scales $\Delta t \gg m^{-1}$.
However, such oscillatory terms must be treated with care.
Suppose, for example, that the full solution of
Eq.~(\ref{psieom1}) consists of a slowly varying contribution
$(\psi_s)$ and a small term that does not vary slowly: $\psi =
\psi_s + \delta \psi e^{2imt}$. Inserting this ansatz back into
Eq.~(\ref{psieom1}), keeping only terms up to linear order in
$\delta \psi$, and retaining terms that vary in time more slowly
than $e^{\pm i mt}$, we obtain
\beq
i \dot{\psi}_s \simeq - \frac{1}{2m} \nabla^2 \psi_s + \frac{
\lambda}{8 m^2} \vert \psi_s \vert^2 \psi_s +
\frac{\lambda}{8m^2} \psi_s \left( \psi_s \delta \psi + 2
\psi_s^* \delta \psi^* \right) + ...
\label{psieom2}
\eeq
We find that the fast-oscillating portion of the solution
$(\delta \psi)$ backreacts on the slowly varying portion
$(\psi_s)$, contributing to the dynamics of $\psi_s$ in a
nontrivial way. (Cf.~Refs.~\cite{memory1,memory2}.) This simple
example illustrates that we must take into account the
fast-oscillating contributions of the full solution to obtain an
effective description of the slowly varying portion.  The removal
of the fast-oscillating contributions is analogous to integrating
out the high-frequency components of a field in a path integral.

We construct a perturbative approach with which to account for
the contribution of the fast-oscillating terms to the time
evolution of the slowly varying portion of the field $\psi$.  In
general there are three small quantities to consider for our
perturbative treatment. For any quantity $F (t, {\bf x})$, we may
consider spatial variations compared to the length-scale
$m^{-1}$; time variation compared to the time-scale $m^{-1}$; and
self-interactions mediated by the coupling constant $\lambda$. We
parameterize the magnitude of spatial and temporal variations as
\beq
\frac{ \nabla^2 F }{ m^2} \sim \epsilon_x F \> , \quad\quad \frac{ \dot{F} }{ m } \sim \epsilon_t F .
\label{epsilonxt}
\eeq
For weakly interacting systems in the nonrelativistic limit, we
assume that $\epsilon_x , \epsilon_t, \lambda \ll 1$. Given our
nonlocal field redefinition in terms of the operator ${\cal P}$,
we may manipulate quantities to arbitrary order in $\epsilon_x$
during intermediate steps and expand the ${\cal P}$ operators to
the desired order at the end. We may then Taylor expand various
quantities in powers of $\epsilon_t$ and $\lambda$ and
iteratively track their effects on the slowly varying portion of
the field. Upon expanding the ${\cal P}$ operators, we may use
the equation of motion (at appropriate, iterative order) to
relate higher-order terms in $\epsilon_t$ to corresponding
higher-order terms in $\epsilon_x$ and $\lambda$. In the end this
yields a controlled expansion in two small parameters,
$\epsilon_x$ and $\lambda$.

We construct the solution for $\psi (t, {\bf x})$ in the form of
an expansion in an infinite series of harmonics,
\beq
\psi (t, {\bf x}) = \sum_{\nu = - \infty}^\infty \psi_\nu (t, {\bf x} ) e^{i \nu mt} ,
\label{psinu}
\eeq
where each $\psi_\nu (t, {\bf x} )$ will be slowly varying on a
time scale of order $m^{-1}$. We label the mode with $\nu = 0$ as
the slowly varying contribution to the field,
\beq
\psi_{\nu = 0} (t, {\bf x} ) \equiv \psi_s (t, {\bf x} ).
\label{psislow}
\eeq
We assume that in the nonrelativistic limit, $\vert \psi - \psi_s
\vert \ll \vert \psi_s \vert$.

We may rewrite the equation of motion for $\psi$ in
Eq.~(\ref{psieom1a}) in the form
\beq
i \dot{\psi} (t, {\bf x} ) = m \left( {\cal P} - 1 \right) \psi
(t, {\bf x}) + \frac{\lambda}{4! \, m^2} \tilde{G} (t, {\bf x} )
,
\label{eomG}
\eeq
where we have defined
\beq
\tilde{G} (t, {\bf x} ) \equiv {\cal P}^{-1/2} e^{imt} \left[ e^{-imt} \, {\cal P}^{-1/2} \psi + e^{imt} \, {\cal P}^{-1/2} \psi^* \right]^3 .
\label{Gdef}
\eeq
We expand $\tilde{G}$ in a series akin to the expansion for
$\psi$:
\beq
\tilde{G} (t, {\bf x}) = \sum_{\nu = - \infty}^\infty \tilde{G}_\nu (t, {\bf x}) e^{i \nu mt} .
\label{psinu2}
\eeq
For a given mode $\nu$, Eq.~(\ref{eomG}) then takes the form
\beq
i \dot{\psi}_\nu - \nu m \psi_\nu = m \left( {\cal P} - 1 \right)
\psi_\nu + \frac{\lambda}{4! \, m^2} \tilde{G}_\nu
\label{eompsinu}
\eeq
with
\beq
\tilde{G}_\nu (t, {\bf x}) = {\cal P}^{-1/2} \sum_{\mu, \mu'} \bigg\{ \Psi_\mu \Psi_{\mu'} \Psi_{2 + \nu - \mu - \mu'} + \Psi^*_\mu \Psi^*_{\mu'} \Psi^*_{4 - \nu - \mu - \mu'} + 3 \Psi_\mu \Psi_{\mu' } \Psi^*_{\mu + \mu' - \nu} + 3 \Psi^*_\mu \Psi^*_{\mu '} \Psi_{\nu - 2 + \mu + \mu'} \bigg\} ,
\label{Gnu}
\eeq
where we have made use of the convenient notation
\beq
\Psi_\nu (t, {\bf x} )\equiv {\cal P}^{-1/2} \psi_\nu  (t, {\bf x} ).
\label{Psinu3}
\eeq
For $\nu=0$, Eq.~(\ref{eompsinu}) gives us the equation obeyed by
$\psi_s(t,{\bf x})$, which defines the low-energy effective field
theory that we seek.  However, to evaluate $\tilde G_\nu$, we
need to (perturbatively) calculate $\psi_\nu(t, {\bf x})$ for all
other values of $\nu$. Before proceeding, we note that for each
$\nu$, the mode $\psi_\nu$ carries a definite charge. In
particular, if we assign charges $Q_\nu = 1 - \nu$ to $\psi_\nu$
and $\bar{Q}_\nu = - Q_\nu$ to $\psi_\nu^*$, then total charge
remains conserved because all terms in Eq.~(\ref{eompsinu}) carry
the same charge. This implies that the mode of interest,
$\psi_s$, carries a conserved charge, and hence we expect the
effective field theory governing $\psi_s$ to obey a global $U(1)$
symmetry, to all orders in the perturbative expansion, in spite
of the fact that the exact theory violates this symmetry.  Below
we will explicitly confirm that particle number conservation is
exact for the leading perturbative corrections.

Eq.~(\ref{eompsinu}) is a first-order differential equation in
time, which in principle defines $\psi_\nu(t,{\bf x})$ up to an
arbitrary function of ${\bf x}$. However, as long as we can
construct a function $\psi_\nu(t,{\bf x})$ which satisfies
Eq.~(\ref{eompsinu}), then the full series of Eq.~(\ref{psinu})
will satisfy the equation of motion, Eq.~(\ref{psieom1a}), so it
will be sufficient for us to construct a particular solution to
Eq.~(\ref{eompsinu}).  To do so, we multiply both sides of
Eq.~(\ref{eompsinu}) by ${\cal P}^{-1/2}$ and rearrange terms to
write
\beq
\Psi_\nu = - \frac{ i}{m} \Gamma_\nu \dot{\Psi}_\nu + \lambda G_\nu ,
\label{PsinuGamma}
\eeq
where we have defined
\beq
\Gamma_\nu \equiv (1 - \nu - {\cal P} )^{-1}
\label{Gammadef}
\eeq
and
\beqn
\begin{split}
G_\nu (t, {\bf x}) &= \frac{ \Gamma_\nu {\cal P}^{-1/2} }{4! \,
m^3} \tilde G_\nu(t, {\bf x}) \\ &= \frac{ \Gamma_\nu {\cal
P}^{-1} }{4! \, m^3}
   \sum_{\mu, \mu'} \bigg\{ \Psi_\mu \Psi_{\mu'} \Psi_{2 + \nu -
   \mu - \mu'} + \Psi^*_\mu \Psi^*_{\mu ' } \Psi^*_{4 - \nu - \mu
- \mu ' } + 3 \Psi_\mu \Psi_{\mu ' } \Psi^*_{\mu + \mu' - \nu}\\
& \qquad\qquad\qquad\quad  + 3 \Psi^*_\mu \Psi^*_{\mu ' } \Psi_{\nu - 2 + \mu + \mu
' }
   \bigg\} .
\label{Gnudef}
\end{split}
\eeqn
Eq.~(\ref{PsinuGamma}) holds to any order in $\epsilon_x$,
$\epsilon_t$, and $\lambda$. We note, however, that the first
term on the right-hand side is suppressed relative to $\Psi_\nu$
by a factor of $\epsilon_t$ and the second term is suppressed
relative to $G_\nu$ by a factor of $\lambda$, so we may treat the
right-hand side as a perturbative source for $\Psi_\nu$.
Therefore we can solve Eq.~\eqref{PsinuGamma} iteratively,
starting with the zeroth-order approximation
\beq
\Psi_\nu^{(0)} (t, {\bf x}) = \begin{cases} 
                        \Psi_s(t, {\bf x}) & \hbox{if } \nu=0\ ,\\
     0 & \hbox{if } \nu \not= 0 \ .
                        \end{cases}
\eeq
We denote by $\Psi_\nu^{(n)}$ the correction to $\Psi_\nu$
obtained at the $n$th iteration, which in our construction will
always be proportional to a total of $n$ powers of the small
quantities $\epsilon_t$ and $\lambda$. We expand the modes with
$\nu \neq 0$ in the series
\beq
\Psi_\nu (t, {\bf x} ) = \sum_{n = 0}^\infty  \Psi_\nu^{(n)} (t, {\bf x} ) \quad {\rm for} \> \nu \neq 0 ,
\label{psinuepsilon}
\eeq
where the terms $\Psi_\nu^{(n)}$ will be determined iteratively,
as described below, and $\Psi_0^{(n)}
\equiv 0$ for $n > 0$, since we are not expanding $\Psi_0 \equiv
\Psi_s$ in a power series.
We also expand
\beq
G_\nu (t, {\bf x} ) = \sum_{n = 0}^\infty G_\nu^{(n)} (t, {\bf
x}),
\label{Gnuepsilon}
\eeq
where $G_\nu^{(n)} (t, {\bf x})$ consists of all terms in
Eq.~(\ref{Gnudef}) which are proportional to the $n$th power of
the small quantities $\epsilon_t$ and $\lambda$. From
Eq.~(\ref{PsinuGamma}), we have at first order
\beq
\Psi_\nu^{(1)} = \lambda G_{\nu}^{(0)} \quad
{\rm for} \>\> \nu \not=0 \ ,
\label{Psinu1}
\eeq
and for higher orders we have
\beq
\Psi_\nu^{(n)} = - \frac{i}{m} \Gamma_\nu \dot{\Psi}_\nu^{(n -
1)} + \lambda G_{\nu}^{(n - 1)} \quad {\rm for} \>\>
\nu \not= 0 \hbox{ and } n > 1 .
\label{Psinun}
\eeq
The first term on the right-hand side is suppressed relative to
$\Psi_\nu^{(n -1)}$ by $\epsilon_t$, and the second term is
suppressed relative to $G_{\nu}^{(n - 1)}$ by $\lambda$.  Thus,
from Eq.~\eqref{Psinun}, we see by induction that
$\Psi_\nu^{(n)}$ will have a total power in the small quantities
$\epsilon_t$ and $\lambda$ of exactly $n$. 

Our aim is to calculate the effective equation of motion for
$\psi_s$, which is given by Eqs.~(\ref{eompsinu}) and
(\ref{Gnudef}) as
\beq
i \dot{\psi}_s = m ({\cal P} - 1) \psi_s + m \lambda
\Gamma_0^{-1} {\cal P}^{1/2} G_0 ,
\label{eompsis1}
\eeq
with $G_0$ expanded to some perturbative order, as in
Eq.~(\ref{Gnuepsilon}). Working to order $n = 1$ we find
\beqn
\begin{split}
G_\nu^{(0)} &= \frac{1}{\lambda} \Psi_\nu^{(1)} = \frac{
\Gamma_\nu {\cal P}^{-1} }{4! \, m^3} \bigg\{ \Psi_s^3
\delta_{\nu, -2} + \Psi_s^{* 3} \delta_{\nu, 4} + 3 \vert \Psi_s
\vert^2 \Psi_s^* \delta_{\nu, 2} + 3 \vert \Psi_s \vert^2 \Psi_s
\delta_{\nu, 0} \bigg\} , \\ G_\nu^{(1)} &= \frac{ 3 \Gamma_\nu
{\cal P}^{-1}}{4! \, m^3} \bigg\{ \Psi_s^2 \Psi_{2 + \nu}^{(1)} +
\Psi_s^{* 2} \Psi_{4 - \nu}^{(1) *} + \Psi_s^2 \Psi_{- \nu}^{(1)
*} + 2 \vert \Psi_s \vert^2 \Psi_\nu^{(1)} + \Psi_s^{* 2}
\Psi_{\nu - 2}^{(1)} + 2 \vert \Psi_s \vert^2 \Psi_{2 - \nu}^{(1)
*} \bigg\} ,
\end{split}
\label{G0G1}
\eeqn
where $\delta_{i,j}$ is the Kronecker delta function and, as
discussed after Eq.~(\ref{psinuepsilon}), we set $\Psi_0^{(n)} =
0$. Using the expressions in Eq.~(\ref{G0G1}), we may expand
Eq.~(\ref{eompsis1}) to order $n = 1$, which yields
\beqn
\begin{split}
i \dot{\psi}_s &= m ({\cal P} - 1) \psi_s + \frac{ \lambda {\cal
P}^{-1/2} }{8 m^2} \vert \Psi_s \vert^2 \Psi_s \\ &\quad\quad +
\frac{ 3 \lambda^2 {\cal P}^{-1/2} }{(4!)^2 m^5} \bigg\{ 3
\Psi_s^2 \Gamma_2 {\cal P}^{-1} (\vert \Psi_s \vert^2 \Psi_s^* )
+ \Psi_s^{* 2} \Gamma_4 {\cal P}^{-1} (\Psi_s^3) \\
&\quad\quad\quad\quad\quad\quad \quad\quad + \Psi_s^{* 2}
\Gamma_{-2} {\cal P}^{-1} (\Psi_s^3 ) + 6 \vert \Psi_s \vert^2
\Gamma_2 {\cal P}^{-1} (\vert \Psi_s \vert^2 \Psi_s ) \bigg\} \\
&\quad\quad + {\cal O} [\lambda^3, \epsilon_t^3, \lambda^2
\epsilon_t , \lambda \epsilon_t^2 ] .
\end{split}
\label{eompsis2}
\eeqn
We may now expand the ${\cal P}$ operators to order $\epsilon_x^2
\sim \nabla^4 / m^4$, which yields

\beqn
\begin{split}
i\dot{\psi}_s &\simeq - \frac{1}{2m} \nabla^2 \psi_s + \frac{
\lambda}{8 m^2} \vert \psi_s \vert^2 \psi_s \\ &\quad\quad -
\frac{1}{8 m^3} \nabla^4 \psi_s + \frac{ \lambda}{32 m^4} \bigg[
\psi_s^2 \nabla^2 \psi_s^* + 2 \vert \psi_s \vert^2 \nabla^2
\psi_s + \nabla^2 \left( \vert \psi_s \vert^2 \psi_s \right)
\bigg] - \frac{17 \lambda^2}{768 m^5} \vert \psi_s \vert^4 \psi_s
\\ &\quad\quad\quad\quad + {\cal O} [ \lambda^3 , \epsilon_t^3 ,
\epsilon_x^3, \lambda^2 \epsilon_t , \lambda^2 \epsilon_x ,
\lambda \epsilon_t^2 , \lambda \epsilon_x^2 , \lambda \epsilon_t
\epsilon_x, \epsilon_t \epsilon_x^2 , \epsilon_t^2 \epsilon_x] .
\end{split}
\label{eompsislow}
\eeqn

To obtain Eq.~(\ref{eompsislow}), we only needed to calculate the
$\Psi_\nu^{(n)}$ up to $n=1$, so it was sufficient to use
Eq.~(\ref{Psinu1}), and the time derivative in Eq.~(\ref{Psinun})
never appeared.  But of course if we continued to the next order,
the time derivative of $\Psi_\nu^{(n-1)}$ would appear.  The
iteration of this procedure will lead to time derivatives higher
than the first, which will give at least the appearance of
introducing new degrees of freedom.  There exists, however, a
well-defined procedure for treating these higher time derivatives
perturbatively, so that no new degrees of freedom are introduced. 
We illustrate these techniques in Appendices
\ref{sec:local_redef} and \ref{sec:Masaki_compare}.  The higher
time derivatives can be removed later, but in the context of the
iteration procedure described here, they can be eliminated at
each iteration.  To do this, start by writing $\Psi_\nu^{(n-1)}$
in terms of $\psi_s$.  For $n=2$, this would be the top line of
Eq.~(\ref{G0G1}). At each stage, we will be able to express
$\Psi_\nu^{(n-1)}$ in a perturbative expansion, with all terms up
to a total power of $n-1$ in the small quantities $\epsilon_t$
and $\lambda$.  Then we can differentiate this expression with
respect to $t$, evaluating $\dot\psi_s$ according to
Eq.~(\ref{eompsis1}).  Even though $\Psi_\nu^{(n-1)}$ has a total
power of $n-1$ in the small quantities $\epsilon_t$ and
$\lambda$, its time derivative can be calculated up to one power
higher, since the time derivative brings an extra factor of
$\epsilon_t$.  Thus the first term on the right-hand side of
Eq.~(\ref{Psinun}) can be evaluated as a term with no time
derivatives, and with a total power of $n$ in the small
quantities $\epsilon_t$ and $\lambda$, which is exactly what is
wanted.

The first line of Eq.~(\ref{eompsislow}) is the usual
Schr\"{o}dinger equation for a self-interacting scalar field in
the nonrelativistic limit, and the second line includes the
lowest-order relativistic corrections. In the second line, the
terms that include the Laplacian operator come from expanding the
operator ${\cal P}$, and the final term comes from the
contribution of the fast-oscillating terms $\psi_\nu^{(1)}$ (for
$\nu \neq 0$).  Given the equation of motion (\ref{eompsislow}),
we can find an effective Lagrangian which produces this equation
of motion:
\beqn 
\begin{split}
{\cal L}_{\rm eff} &= \frac{i}{2} \left( \dot{\psi}_s \psi_s^* -
\psi_s \dot{\psi}_s^* \right) - \frac{1}{2m} \nabla \psi_s \nabla
\psi_s^* - \frac{ \lambda}{16 m^2} \vert \psi_s \vert^4 \\
&\quad\quad + \frac{1}{ 8 m^3} \nabla^2 \psi_s \nabla^2 \psi_s^*
- \frac{ \lambda}{32 m^4} \vert \psi_s \vert^2 \left( \psi_s^*
\nabla^2 \psi_s + \psi_s \nabla^2 \psi_s^* \right) + \frac{17
\lambda^2}{9 \times 2^8 \, m^5} \vert \psi_s \vert^6 .
\end{split}
\label{Lpsislow}
\eeqn

We note that a single self-interaction term in the relativistic
Lagrangian (in this case, $\lambda \phi^4 / 4!$) yields more than
one interaction term in the nonrelativistic effective theory. In
particular, the effective Lagrangian for $\psi_s$ in
Eq.~(\ref{Lpsislow}) includes interaction terms that yield both
$2 \rightarrow 2$ scattering and $3 \rightarrow 3$ scattering;
including more terms in the iterative expansion for $\psi_\nu$
(with $\nu \neq 0$) would yield operators in ${\cal L}_{\rm eff}$
for each $n \rightarrow n$ scattering, for $n > 1$. On the other
hand, ${\cal L}_{\rm eff}$ obeys a global $U(1)$ symmetry, a
feature which holds to all orders in the perturbative expansion,
as discussed after Eq.~(\ref{Psinu3}).  Thus, particle number is
conserved to all orders in the nonrelativistic effective field
theory. A process that may occur in the relativistic theory, such
as $4
\rightarrow 2$ scattering in which four low-energy particles
annihilate to produce two relativistic ones, involves energies $E
> m$ and hence lies beyond the range of validity of ${\cal
L}_{\rm eff}$ in Eq.~(\ref{Lpsislow}). 

We have obtained the effective theory in Eq.~(\ref{Lpsislow})
starting from the nonlocal field redefinition of
Eqs.~\eqref{phipsi1}. Our procedure was relatively straightforward
because the equation of motion in the free-field limit does not
contain fast-oscillating factors, and all of the
spatial-derivative operators may be manipulated in terms of the
nonlocal operator ${\cal P}$. Whereas these steps simplify the
calculation when working with the nonlocal field redefinition of
Eqs.~(\ref{phipsi1}), one may follow similar steps to derive an
equivalent effective field theory starting from the local field
redefinition of Eqs.~(\ref{phitrad}), as we demonstrate in
Appendix \ref{sec:local_redef}. 

We may apply Eqs.~(\ref{eompsislow}) and (\ref{Lpsislow}) to the
case of QCD axions in the nonrelativistic limit, and compare our
effective description with other recent treatments
\cite{GHPW15,Braaten16}. We begin with the usual relativistic
potential,
\beq
V (\phi) = \Lambda^4 \left[ 1 - \cos (\phi / f_a ) \right] ,
\label{Vaxion}
\eeq
where $\Lambda \sim 0.1$ GeV is associated with the QCD scale,
and $f_a$ sets the scale for Peccei--Quinn symmetry breaking; we
expect $f_a \sim 10^{11} - 10^{12}$ GeV for the typical
Peccei--Quinn model \cite{axionCP1,axionCP2,axionCP3,GHPW15}. For
field values well below $f_a$, we may expand $V (\phi)$ as
\beq
V (\phi) = \frac{1}{2} m^2 \phi^2 + \frac{\lambda}{4!} \phi^4 +
{\cal O} \left[ (\phi / f_a)^6 \right] ,
\label{Vaxion2}
\eeq
where $m = \Lambda^2 / f_a $ and $\lambda = - \Lambda^4 / f_a^4 <
0$. Given expected values for $\Lambda$ and $f_a$, these
correspond to $m \sim 10^{-4}$ -- $10^{-5}$ eV and $\vert \lambda
\vert \sim 10^{-48}$ -- $10^{-52}$.

Remarkably, the final term in Eq.~(\ref{eompsislow}),
proportional to $\lambda^2$, exactly matches the corresponding
term found in Ref.~\cite{Braaten16}, following quite a different
procedure. (In Ref.~\cite{Braaten16}, the authors fixed the
coefficients in a series expansion for the low-energy effective
potential by calculating scattering amplitudes for various
$n$-body scattering processes in the full, relativistic theory,
and then took the low-energy limit of those amplitudes.) On the
other hand, the analysis in Ref.~\cite{Braaten16} does not
capture the relativistic corrections present in
Eq.~(\ref{eompsislow}), proportional to $\nabla^4$ and $\lambda
\nabla^2$. We may compare magnitudes for the $\lambda \nabla^2$
and $\lambda^2$ terms:
\beq
\left( \frac{ 17 \lambda^2 \vert \psi_s \vert^4 \psi_s}{768 m^5} \right) \left(\frac{ 32 m^2}{  \lambda   \vert \psi_s \vert^2 \nabla^2 \psi_s }\right) = {\cal O} (1) \left( \frac{ \lambda \vert \psi_s \vert^2 \psi_s }{8 m^2} \right) \left( \frac{ 2m}{\nabla^2 \psi_s } \right) \sim \frac{\lambda {\cal N} }{ m^3 v^2}  ,
\label{termcompare}
\eeq
where ${\cal N} = \vert \psi_s \vert^2$ is the axion number
density, and we have used $\nabla^2 / m^2 \sim v^2$. Thus we see
that whenever the contributions arising from kinetic and
potential energy to the zeroth-order equation of motion are of
comparable magnitude to each other --- that is, the two terms on
the right-hand side of the top line of Eq.~(\ref{eompsislow}) ---
then the leading-order correction terms, on the second line of
Eq.~(\ref{eompsislow}), will also be of comparable magnitude to
each other. In such situations, it is inconsistent to retain (for
example) the $\lambda^2$ corrections while neglecting the
$\lambda \nabla^2$ corrections. Put another way, one must retain
each of the corrections on the second line of
Eq.~(\ref{eompsislow}) for situations in which the axions have
virialized, with comparable (time-averaged) values of kinetic and
potential energy \cite{Virialfn}. We further note that in
Ref.~\cite{Braaten16}, the term $m \vert \psi_s \vert^2$ appears
in their low-energy Hamiltonian (Eqs.~(26) and (27) of their
paper), whereas we find no such term in ours. As we show in
Appendix \ref{sec:canonical} (see Eqs.~\eqref{canonical} and
\eqref{Delta_H}), the $m \vert \psi_s\vert^2$ term is exactly
canceled by a compensating term arising from the canonical
transformation.

We may also compare our results in Eq.~(\ref{eompsislow}) with
the recent calculation in Ref.~\cite{MasakiOscillons}.
Superficially our equations of motion appear to disagree at order
$\lambda^2$. However, as demonstrated in Appendix
\ref{sec:Masaki_compare}, our results are completely consistent
(at least to this perturbative order), as can be shown by
performing a nonlinear field redefinition.

Finally, we note that if one considers corrections from modes
$\psi_\nu^{(n)}$ with $\nu \neq 0$ and $n \geq 2$, then one will
generically find higher-order time-derivative terms in the
effective equation of motion for $\psi_s$, as discussed after
Eq.~(\ref{eompsislow}). This is because the mode functions for
fast-oscillating terms $\psi_\nu^{(j)}$ depend, in general, on
both $\psi_\nu^{(i)}$ and $\dot{\psi}_\nu^{(i)}$ for $i < j$.
Such higher-order time derivatives do not introduce unphysical
degrees of freedom --- after all, the full, relativistic theory
remains well-behaved. Rather, such higher-order time derivatives
typically arise in perturbative expansions for low-energy
effective field theories, and may be removed systematically by
applying the equations of motion at an appropriate perturbative
order \cite{GrosseKnetter94}. This procedure corresponds to
excluding certain solutions of the low-energy effective theory
that could not have been considered perturbations around the
zeroth-order equations of motion, and hence remains consistent
with the spirit of constructing effective field theories
\cite{GrosseKnetter94,Georgi91,Arzt95}.

\section{Conclusions}
\label{sec:conclusions}

In this paper we have developed a self-consistent framework for
obtaining an effective field theory to describe the
nonrelativistic limit of a relativistic field theory. The
lowest-order corrections to the ultra-nonrelativistic limit arise
both from expanding the kinetic energy as well as from
incorporating the backreaction from fast-oscillating terms on the
dominant, slowly varying portion of the field. Our results are
largely consistent with the complementary analyses of
Refs.~\cite{MasakiOscillons} and \cite{Braaten16}.  Our approach
is perhaps simpler, working directly with the equations of motion
in the nonrelativistic limit.  In addition, our approach
incorporates nontrivial relativistic corrections in a systematic
way, which had been neglected in previous analyses. These
additional terms may be comparable in magnitude to the other,
known terms in various physical situations of interest.

Rather than begin with the usual relation of Eqs.~(\ref{phitrad})
between the real, relativistic scalar field and the complex,
nonrelativistic field, we introduce a fairly simple, nonlocal
field redefinition, as in Eqs.~(\ref{phipsi1}). This new field
redefinition considerably simplifies the treatment for free
fields, and enables us to construct an iterative, perturbative
procedure in the presence of interactions. 

Other than imposing the standard commutation relations, our
treatment relies on manipulating classical fields.  Even so, we
believe our formalism captures the relevant dynamics of the
low-energy effective quantum field theory --- including
higher-order corrections in the coupling constant $\lambda$. 
The tree-level diagrams of the quantum theory directly reflect 
the equations of motion of the classical theory, and we would 
expect that any low-energy effective Lagrangian that produces 
the correct tree-level diagrams will also produce the correct 
loop diagrams. In our formalism we
find that particle number is conserved to all orders in the
nonrelativistic, perturbative description; an interesting
question is how best to incorporate particle-number-violating
processes in a low-energy effective description, in a
self-consistent way. (Cf.~Refs.~\cite{MasakiOscillons} and \cite{BraatenStars16}.)

We have focused on the dynamics of a single real,
self-interacting scalar field. Applications of interest, which we
intend to explore in future work, include the behavior of axion
dark matter in galactic halos (building on
Refs.~\cite{GHPW15,HertzbergNamjooCore} and references therein),
as well as in hypothetical axion stars
\cite{BraatenStars16,axionstar1,axionstar2,axionstar3,axionstar4,axionstar5,Schiappacasse:2017ham,Visinelli17},
for which one must supplement our treatment here to incorporate
gravitational interactions. It would also be interesting to
extend the approach developed here to other types of fields, such
as the Dirac field. 

{\it Note added:} As we were finalizing revisions to this paper, Ref.~\cite{BraatenHigherOrder} appeared, which reaches similar conclusions regarding relations among various approaches to the low-energy effective descriptions of a given relativistic quantum field theory.

\appendix

\section{Canonical Derivation of the Field Redefinition}
\label{sec:canonical}

Our calculations have been based on the field redefinition of
Eqs.~(\ref{phipsi1}), and their inverse given by
Eq.~\eqref{psieq}. In Eq.~\eqref{Lrelpsi} we constructed a
Lagrangian that gives the correct equation of motion for $\psi$,
but in this appendix we show how to construct the field
redefinition as a canonical transformation.  Canonical
transformations are guaranteed to preserve the Poisson bracket
relations, which become the commutators upon quantization.

We start by adopting the real part of $\psi$ as the new canonical
field $\psi_c$, but we allow a normalization factor ${\cal C}_1$
that can later be adjusted for convenience:
\beq
\psi_c = {\cal C}_1 \psi_R ,
\label{N_1}
\eeq
where $\psi \equiv \psi_R + i \psi_I$.  The canonical momentum
conjugate to $\psi_c$ will be called $\pi_c$, which we expect to
be proportional to $\psi_I$:
\beq
\pi_c = {\cal C}_2 \psi_I .
\label{N_2}
\eeq
$\phi$ and $\pi$ will be used to denote the original relativistic
field and its canonical conjugate.  To describe the canonical
transformation, we adapt the discrete-variable formalism of
Goldstein, Poole, and Safko~\cite{Goldstein3} to the case of
continuous fields. The canonical transformation is then defined
in terms of a generating functional $F_2[\phi, \pi_c, t]$, and
the canonical transformation is described by
\beq
\pi(x) = \frac{\delta F_2}{\delta \phi(x)} \ , \qquad \psi_c(x) =
\frac{\delta F_2}{\delta \pi_c(x)}\ , \qquad H_{\rm new} = H +
\frac{\partial F_2}{\partial t} ,
\label{canonical}
\eeq
where the partial derivatives in Ref.~\cite{Goldstein3} have been
replaced by functional derivatives. To try to match these
equations, we first rewrite the transformation in terms of the
real and imaginary parts of $\psi$:
\beqn
\begin{split}
\phi(t,{\bf x}) &= \sqrt{\frac{2}{m}}{\cal P}^{-1/2} \left[ \cos
(m t) \psi_R(t,{\bf x}) + \sin (m t) \psi_I(t,{\bf x}) \right],
\\
\pi(t,{\bf x}) &= \sqrt{2 m} {\cal P}^{1/2} \left[ \cos (m t)
\psi_I(t,{\bf x}) - \sin (m t) \psi_R(t,{\bf x}) \right]. 
\end{split}
\label{phipsi2}
\eeqn
We can then algebraically manipulate these equations to find
expressions for $\pi(x)$ and $\psi_c(x)$ in terms of $\phi$ and
$\pi_c$, to compare with Eqs.~\eqref{canonical}:
\beqn
\begin{split}
\psi_c(t,{\bf x}) &= {\cal C}_1 \left[
\sqrt{\frac{m}{2}} {\cal P}^{1/2} \sec (m t) \phi(t,{\bf x}) -
\frac{1}{{\cal C}_2} \tan (m t)
\pi_c(t,{\bf x}) \right] , \\
\pi(t,{\bf x}) &= \frac{1}{{\cal
C}_2} \sqrt{2 m} {\cal P}^{1/2} \sec (m t)
\pi_c(t,{\bf x}) - m {\cal P} \tan (m t) \phi(t,{\bf x}) .
\label{mixedsol}
\end{split}
\eeqn
To be as clear as possible, we express ${\cal P}$ and ${\cal
P}^{1/2}$ as integral operators,
\beqn
\begin{split}
({\cal P} f)({\bf x}) &\equiv \int d^3 y \, {\cal P}({\bf x},
{\bf
     y}) f({\bf y)} ,\\ ({\cal P}^{1/2} f)({\bf x}) &\equiv \int
d^3 y \, {\cal P}^{1/2}({\bf
     x}, {\bf y}) f({\bf y)} ,
\end{split}
\eeqn
where ${\cal P}({\bf x},{\bf y})$ and ${\cal P}^{1/2}({\bf
x},{\bf y})$ are both symmetric, and ${\cal P}^{1/2}$ is the
operator square root of ${\cal P}$, defined by
\beq
\int d^3 z \, {\cal P}^{1/2}({\bf x}, {\bf z}) {\cal P}^{1/2}({\bf
     z}, {\bf y}) = {\cal P} ({\bf x},{\bf y}) .
\eeq
It can then be seen that it is possible to find $F_2$ by
functionally integrating the first two of Eqs.~\eqref{canonical},
where the consistency of the term involving both $\phi$ and
$\pi_c$ requires that ${\cal C}_1 {\cal C}_2 = 2$.  We choose
${\cal C}_1 = {\cal C}_2 = \sqrt{2}$, so that
\beq
\psi(t,{\bf x}) = \frac{1}{\sqrt{2}} \Bigl(
\psi_c(t,{\bf x}) + i \pi_c(t,{\bf x}) \Bigr) ,
\label{psi_c-pi_c2}
\eeq
in agreement with Eq.~\eqref{psi_c-pi_c}. $F_2$ is then given by
\beqn
\begin{split}
F_2[\phi, \pi_c, t] &= \sqrt{m} \sec (m t) \int d^3 x \, d^3 y \,
\phi(t,{\bf x}) {\cal P}^{1/2} ({\bf x}, {\bf y}) \pi_c(t,{\bf
y}) \\ & \qquad - \frac{1}{2} m \tan (m t) \int d^3 x \, d^3 y \,
\phi(t,{\bf x}) {\cal P} ({\bf x},{\bf y}) \phi(t,{\bf y}) -
\frac{1}{2} \tan ( m t) \int d^3 x \, \pi_c^2(t,{\bf x}) .
\end{split}
\label{F2}
\eeqn
To calculate the change in the Hamiltonian from the third of
Eqs.~\eqref{canonical}, one differentiates Eq.~\eqref{F2} with
respect to $t$ and then replaces $\phi(t,{\bf x})$ by using
\beq
\left({\cal P}^{1/2} \phi\right)(t,{\bf x}) = \frac{1}{\sqrt{m}}
\left[ \cos (m t) \psi_c(t,{\bf x}) + \sin (m t) \pi_c(t,{\bf x})
\right] ,
\eeq
which follows from Eqs.~\eqref{phipsi2}, along with
Eqs.~\eqref{N_1} and \eqref{N_2}.  The result is
\beq
\label{Delta_H}
\Delta H = \frac{\partial F_2}{\partial t} = - \frac{1}{2} m \int
d^3 x \, \left\{ \psi_c^2 (t,{\bf x}) + \pi_c^2 (t,{\bf x})
\right\} .
\eeq
To find the new Hamiltonian, we must also express the original
Hamiltonian, given by Eq.~\eqref{Hrel}, in terms of the new
variables.  Note that
\beq
\frac{1}{2} \int d^3 x \left\{ m^2 \phi^2 + (\nabla \phi)^2 \right\}
= \frac{1}{2} \int d^3 x \left\{m^2 \phi^2 - \phi \nabla^2 \phi
\right\} = \frac{m^2}{2} \int d^3 x  \, \phi {\cal P}^2 \phi .
\eeq
It is then straightforward to show that the free part of the
original Hamiltonian can be written as
\beq
H_{\rm free} = \frac{1}{2} m \int d^3 x \, \left\{ \pi_c {\cal P}
\pi_c + \psi_c {\cal P} \psi_c \right\} .
\eeq
The full new Hamiltonian, as prescribed by the third of
Eqs.~\eqref{canonical}, is then given by
\beq
H_{\rm new} = \int d^3 x \, {\cal H}_{\rm new} \ , \qquad {\cal
H}_{\rm new} = m \psi^* ({\cal P} - 1 ) \psi +\frac{\lambda}{96
m^2} \left[ e^{- i m t} \psi + e^{i m t} \psi^*\right]^4 ,
\eeq
where $\psi$ is given by Eq.~\eqref{psi_c-pi_c2}.  Note that this
result is in complete agreement with Eq.~\eqref{Hamiltonian}.

\section{Local vs. nonlocal field redefinition}
\label{sec:local_redef}

In order to obtain the effective field theory for the
nonrelativistic field we have used the nonlocal field
redefinition of Eqs.~(\ref{phipsi1}), in contrast to the local
redefinition of Eqs.~(\ref{phitrad}), which is more typically
found in the literature. As discussed in
Secs.~\ref{sec:fieldredef} and \ref{sec:nonrelEFT}, the nonlocal
field redefinition makes the computations easier, although it is
not fundamentally necessary. In this appendix, we begin with the
local field redefinition of Eqs.~(\ref{phitrad}) and obtain the
same low-energy effective theory at order $n = 1$. As a separate,
nontrivial consistency check, we compute the effective field
theory for free fields up to order $\epsilon_x^5 \sim (k/m)^{10}$
(in Fourier space) and again find results compatible with those
obtained with the nonlocal field redefinition, even though, {\it
a priori}, the results appear rather different. To demonstrate
the equivalence, we remove higher-order time-derivative operators
that appear in the EFT in favor of spatial operators, using the
equations of motion. 

\subsection{Effective field theory through $n=1$}
\label{sec:local_EFT}

We begin with a local field redefinition, in which the
relativistic field $\phi$ and the nonrelativistic field $\chi$
are related by
\beqn
\begin{split}
\phi (t, {\bf x}) &= \frac{1}{ \sqrt{ 2m} } \left[ e^{-imt} \chi (t, {\bf x} ) + e^{imt} \chi^* (t, {\bf x} ) \right] , \\
\pi (t, {\bf x} ) &= - i \sqrt{ \frac{ m}{2} } \,  \left[ e^{-imt} \chi (t, {\bf x} ) - e^{imt} \chi^* (t, {\bf x} ) \right] .
\end{split}
\label{phipsi_loc}
\eeqn
By comparing these equations with Eqs.~(\ref{phipsi1}), one can
see that $\chi$ is related to the field $\psi$ that we used in
the main body of this article by the field redefinition
\begin{subequations}
\begin{align}
\chi &= {1 \over 2} \left({\cal P}^{-1/2} + {\cal P}^{1/2}\right) \psi +
{1 \over 2} e^{2 i m t} \left({\cal P}^{-1/2} - {\cal P}^{1/2}
\right) \psi^* \ ,
    \label{psichidefa} \\
\psi &= {1 \over 2} \left({\cal P}^{1/2} + {\cal P}^{-1/2}\right) \chi +
{1 \over 2} e^{2 i m t} \left({\cal P}^{1/2} - {\cal
P}^{-1/2}\right)
    \chi^* \ .
\end{align}
\label{psichidef}
\end{subequations}

The equation of motion for $\chi$ takes the form
\beq
i \dot{\chi} (t, {\bf x} ) = -\dfrac{1}{2m} \nabla^2 \chi -
\frac{e^{2imt} }{2m} \nabla^2 \chi^* + \frac{\lambda}{4! \, m^2}
{\tilde G}^{\text{loc}} (t, {\bf x} ) ,
\label{eomG_loc}
\eeq
where we have defined
\beq
{\tilde G}^{\text{loc}} (t, {\bf x} ) \equiv e^{imt} \left[
e^{-imt} \, \chi + e^{imt} \, \chi^* \right]^3 ,
\label{Gdef_loc}
\eeq
which matches to the similar object, ${\tilde G}$, defined in Eq.
\eqref{Gdef}, in the limit ${\cal P} \to 1$.  Note from
Eq.~(\ref{eomG_loc}) that the global $U(1)$ symmetry is already
broken at the free-field level, obscuring the conservation of
particle number in the nonrelativistic limit.  As before, we
decompose the field into different modes by
\beq
\chi (t, {\bf x}) = \sum_{\nu = - \infty}^\infty \chi_\nu (t, {\bf x} ) e^{i \nu mt} ,
\label{psinu1}
\eeq
and we define the mode with $\nu = 0$ as the slowly varying
portion of the field (in whose evolution we are interested),
\beq
\chi_{\nu = 0} (t, {\bf x} ) \equiv \chi_s (t, {\bf x} ) .
\label{psislow_loc}
\eeq
We then expand ${\tilde G}^{\text{loc}}$ in a series similar to
$\chi$
\beq
\tilde{G} ^{\text{loc}}(t, {\bf x}) = \sum_{\nu = - \infty}^\infty {\tilde G}_\nu^{\text{loc}} (t, {\bf x}) \, e^{i \nu mt} .
\label{psinu_loc}
\eeq
For a given mode $\nu$, Eq.~(\ref{eomG_loc}) then takes the form
\beq
i \dot{\chi}_\nu - \nu m \chi_\nu = -\dfrac{1}{2m} \nabla^2
\chi_\nu -\dfrac{1}{2m} \nabla^2 \chi_{2-\nu}^* +
\frac{\lambda}{4! \, m^2} {\tilde G}_\nu^{\text{loc}} ,
\label{eompsinu_loc}
\eeq
with
\beq
{\tilde G}_\nu^{\text{loc}} (t, {\bf x}) = \sum_{\mu, \mu'}
\bigg\{ \chi_\mu \chi_{\mu'} \chi_{2 + \nu - \mu - \mu'} +
\chi^*_\mu \chi^*_{\mu'} \chi^*_{4 - \nu - \mu - \mu'} + 3
\chi_\mu \chi_{\mu' } \chi^*_{\mu + \mu' - \nu} + 3 \chi^*_\mu
\chi^*_{\mu '} \chi_{\nu - 2 + \mu + \mu'} \bigg\} .
\label{Gnu_loc}
\eeq

For the slowly varying mode, $\chi_s$, Eqs.~(\ref{eompsinu_loc})
and (\ref{Gnu_loc}) yield
\beqn
i \dot{\chi}_s = -\dfrac{1}{2m} \nabla^2 \chi_s -\dfrac{1}{2m}
\nabla^2 \chi_{2}^* +\frac{\lambda }{4!\, m^2 } {\tilde
G}_0^{\text{loc}} ,
\label{eompsislow1_loc}
\eeqn
where
\ba 
{\tilde G}_0^{\text{loc}} =
  \sum_{\mu , \mu' } \bigg\{ \chi_\mu \chi_{\mu '} \chi_{2 - \mu - \mu'} + \chi^*_\mu \chi^*_{\mu ' } \chi^*_{4 - \mu - \mu ' } + 3 \chi_\mu \chi_{\mu ' } \chi^*_{\mu + \mu '} + 3 \chi^*_\mu \chi^*_{\mu ' } \chi_{\mu + \mu ' - 2 } \bigg \} .
  \label{G0loc}
\ea 
Eqs.~(\ref{eompsinu_loc}) and (\ref{eompsislow1_loc}) reveal that
even at linear order in the modes $\chi_\nu$, different modes
couple to each other. This is one source of complication in the
local field redefinition approach. The other difficulty is that
powers of the spatial Laplacian appear at each order of the
iteration, rather than having them all contained in the operator
${\cal P}$.  As a result, we must expand in powers of all three
small quantities ($\epsilon_t$, $\epsilon_x$, and $\lambda$) from
the beginning of our iterative computations. This is in contrast
with the calculation in Sec.~\ref{sec:nonrelEFT}, in which we
used the nonlocal field redefinition, which enabled us to conduct
most of the iterative calculation, until the final step, in terms
of just two small parameters ($\epsilon_t$ and $\lambda$).  Apart
from these technical difficulties, however, nothing prevents us
from obtaining an effective field theory for $\chi_s$ using
Eqs.~(\ref{eompsinu_loc})--(\ref{G0loc}).

It is instructive to rewrite \eqref{eompsinu_loc} as
\beq
  \nu m \chi_\nu  = i \dot{\chi}_\nu + \dfrac{1}{2m} \nabla^2 \chi_\nu + \dfrac{1}{2m} \nabla^2 \chi_{2-\nu}^* - \frac{\lambda}{4! \, m^2} {\tilde G}_\nu^{\text{loc}} ,
\label{eompsinu_locv2}
\eeq
where, in contrast with the nonlocal case in
Eq.~(\ref{PsinuGamma}), we have kept the prefactors on the
left-hand side to avoid the apparent divergence for $\nu=0$. The
right-hand side of Eq.~(\ref{eompsinu_locv2}) can be thought of
as a perturbative source for $\chi_\nu$: the first term is
suppressed relative to $\chi_\nu$ by $\epsilon_t$, the second and
third terms are suppressed relative to $\chi_\nu$ and
$\chi_{2-\nu}$ by $\epsilon_x$, and the last term is suppressed
relative to ${\tilde G}_\nu^{\text{loc}}$ by $\lambda$. 
Therefore, as noted above, our expansion in this case is in
powers of all three small quantities. 

We next expand the modes with $\nu \neq 0$:
\beq
\chi_\nu (t, {\bf x} ) = \sum_{n = 1}^\infty  \chi_\nu^{(n)} (t, {\bf x} ) \quad {\rm for} \> \nu \neq 0 ,
\label{psinuepsilon_loc}
\eeq
where the terms $\chi_\nu^{(n)}$ are to be determined by
iterative approximation, as described below.  We also expand
${\tilde G}_\nu^{\text{loc}}$:
\beq
{\tilde G}_\nu^{\text{loc}} (t, {\bf x} ) = \sum_{n = 0}^\infty
{\tilde G}_\nu^{\text{loc}^{(n)}} (t, {\bf x}),
\label{Gnuepsilon_loc}
\eeq
where ${\tilde G}_\nu^{\text{loc}^{(n)}} (t, {\bf x})$ consists
of all terms in Eq.~(\ref{G0loc}) which are proportional to a
total of $n$ powers of the small quantities $\epsilon_t$,
$\epsilon_x$, and $\lambda$. Note that we decompose $\chi_\nu$
only for $\nu \neq 0$, and, as in Sec.~\ref{sec:nonrelEFT}, we
set $\chi_0^{(0)} \equiv \chi_s$, $\chi_0^{(n)} = 0$ for $n > 0$,
and $\chi_\nu^{(0)} = 0$ for $\nu \not= 0$. Expanding
\eqref{eompsinu_locv2} yields
\beqn
\begin{split}
  \nu m \chi_\nu^{(1)}  &= \dfrac{1}{2m} \nabla^2 \chi_{s}^* \, \delta_{2,\nu} - \frac{\lambda}{4! \, m^2} {\tilde G}_\nu^{\text{loc}^{(0)} } \quad {\text{for} \,\,\, \nu \not=0}\ ,
  \\ 
      \nu m \chi_\nu^{(n)}  &= i \dot{\chi}_\nu^{(n-1)} + \dfrac{1}{2m} \nabla^2\chi_\nu^{(n-1)} + \dfrac{1}{2m} \nabla^2 \chi_{2-\nu}^{{(n-1)}*} - \frac{\lambda}{4! \, m^2} {\tilde G}_\nu^{\text{loc}^{(n-1)}}  \quad {\text{for}} \,  \, \, \nu \not= 0 \hbox{ and } n \geq 2.
\end{split}
\label{eompsinu_locn}
\eeqn
From this iterative expansion one can obtain the effective
equation for $\chi_s$:
\beqn
i \dot{\chi}_s = -\dfrac{1}{2m} \nabla^2 \chi_s -\dfrac{1}{2m}
\nabla^2 \sum_{n=1}^\infty \chi_{2}^{(n)*} + \frac{\lambda }{4!\,
m^2 } \sum_{n=0}^\infty {\tilde G}_0^{\text{loc}^{(n)}} .
\label{eompsislow1_loc_it}
\eeqn
As in Sec.~\ref{sec:nonrelEFT}, we only compute the leading-order
corrections to the Schr\"{o}dinger equation for $\chi_s$.  From
Eq.~(\ref{Gnu_loc}), we have
\beq
\tilde G_\nu^{\text{loc}^{(0)}} = \chi_s^3 \delta_{\nu,-2} + \chi_s^{* 3}
  \delta_{\nu,4} + 3 \vert \chi_s \vert^2 \chi_s^* \delta_{\nu,2}
+ 3 \vert \chi_s \vert^2 \chi_s \delta_{\nu,0} \ ,
\eeq
and then to order $n = 1$, we find
\beq
\chi_\nu^{(1)} =\dfrac{1}{4m^2} \nabla^2 \chi_s^* \, \delta_{\nu,2}+
 \frac{ \lambda}{2 \times 4! \, m^3}  \left[ \chi_s^3 \,
\delta_{\nu ,- 2} - \frac{1}{2} \chi_s^{* 3} \, \delta_{\nu,4} -
3 \vert \chi_s \vert^2 \chi_s^* \, \delta_{\nu , 2} \right]
\quad \text{for}\ \nu \not= 0 \ .
\label{psinu1_loc}
\eeq
Again from Eq.~(\ref{Gnu_loc}), we find
\beq
\tilde G_\nu^{\text{loc}^{(1)}} = 3 \left[ \chi_s^2 \chi_{2 + \nu}^{(1)}
+ \chi_s^{*2} \chi_{4-\nu}^{(1)*} + \chi_s^2
   \chi_{-\nu}^{(1)*}  + 2 \vert \chi_s \vert^2 \chi_\nu^{(1)} +
   \chi_s^{* 2} \chi_{\nu-2}^{(1)} + 2 \vert \chi_s \vert^2
   \chi_{2-\nu}^{(1)*} \right] \ ,
\eeq
so
\beqn
\begin{split}
i \dot{\chi}_s &= -\dfrac{1}{2m} \nabla^2 \chi_s + \frac{ 3
\lambda }{4! \, m^2} \vert \chi_s \vert^2 \chi_s \\ &\quad\quad
-\dfrac{1}{2m} \nabla^2 \chi_2^{(1)*} + \frac{3 \lambda }{4! \,
m^2} \left[ \chi_s^2 \chi_2^{(1)} + \chi_s^{* 2} \chi_4^{(1) *} +
\chi_s^{* 2} \chi_{-2}^{(1)} + 2 \vert \chi_s \vert^2 \chi_2^{(1)
*} \right] \\ &\quad\quad + {\cal O} \left( \epsilon^2 \right) .
\end{split}
\label{eompsislow2_loc}
\eeqn
For convenience, we have introduced the shorthand
notation
\beqn
\begin{split}
{\cal O} ( \epsilon^2) &\equiv {\cal O} \left[
\epsilon_x^2,\epsilon_t^2, \lambda^2, \epsilon_x \epsilon_t,
\epsilon_x \lambda , \epsilon_t \lambda \right] , \\ {\cal O}
(\epsilon^3 ) &\equiv {\cal O} \left[ \epsilon_x^3, \epsilon_t^3
, \lambda^3, \epsilon_x^2 \epsilon_t , \epsilon_x^2 \lambda ,
\epsilon_x \epsilon_t^2 , \epsilon_x \lambda^2 , \epsilon_x
\epsilon_t \lambda , \epsilon_t^2 \lambda , \epsilon_t \lambda^2
\right] .
\end{split}
\label{Orderepsilon}
\eeqn 
Next we substitute $\chi_\nu^{(1)}$ from Eq.~(\ref{psinu1_loc})
into Eq.~(\ref{eompsislow2_loc}), which yields
\beqn
\begin{split}
i\dot{\chi}_s &\simeq - \frac{1}{2m} \nabla^2 \chi_s + \frac{
\lambda}{8 m^2} \vert \chi_s \vert^2 \chi_s \\ &\quad\quad -
\frac{1}{8 m^3} \nabla^4 \chi_s + \frac{ \lambda}{32 m^4} \bigg[
\chi_s^2 \nabla^2 \chi_s^* + 2 \vert \chi_s \vert^2 \nabla^2
\chi_s + \nabla^2 \left( \vert \chi_s \vert^2 \chi_s \right)
\bigg] - \frac{17 \lambda^2}{768 m^5} \vert \chi_s \vert^4 \chi_s
\\
&\quad\quad\quad + {\cal O} (\epsilon^3 ) .
\end{split}
\label{eompsislow_loc}
\eeqn
Eq.~(\ref{eompsislow_loc}) for $\chi_s$ exactly matches
Eq.~(\ref{eompsislow}) for $\psi_s$, which is related to the
relativistic field $\phi$ via the nonlocal field redefinition of
Eqs.~(\ref{phipsi1}). 

Since $\chi$ and $\psi$ are related to each other by the
nontrivial field redefinition of Eqs.~(\ref{psichidef}), there
was no guarantee that $\chi_s$ and $\psi_s$ should obey the same
equation.  The two field theories must be equivalent, since they
are both equivalent to the low-energy effective field theory
derived from the relativistic $\phi^4$ theory, but $\chi_s$ and
$\psi_s$ could be related by a field redefinition that causes
them to have different equations of motion.  We have found,
however, that through ${\cal O}(\epsilon^2)$, the equations for
$\chi_s$ and $\psi_s$ are identical.  We do not know if this
relation will continue to hold at higher orders.  However, for
the special case of the free field theory, $\lambda=0$, the
relation between $\chi_s$ and $\psi_s$ is simple.  For the free
field theory, $\psi$ itself is slowly varying, so $\psi_s =
\psi$.  Then in the equation (\ref{psichidefa}), which expresses
$\chi$ in terms of $\psi$, the first term is purely slowly
varying, while the second term is the product of $e^{2 i m t}$
and a slowly varying function.  Thus, the first term is the
slowly varying part of $\chi$:
\beq
   \chi_s = {1 \over 2} \left({\cal P}^{-1/2} + {\cal P}^{1/2}
     \right) \psi_s \ . 
   \label{chispsis}
\eeq
Since $\left({\cal P}^{-1/2} + {\cal P}^{1/2} \right)$ commutes
with the differential operators in the equation of motion,
$\chi_s$ obeys the same equation of motion as $\phi_s$,
\beq
i \dot \chi_s = m ( {\cal P} - 1) \chi_s \ .
\eeq 

The iterative procedure used above can be continued to obtain
higher- and higher-order corrections to the Schr\"{o}dinger
equation. Generically, higher-order terms contain higher-order
time-derivative operators, as expected in an EFT framework, and
as was discussed after Eq.~(\ref{eompsislow}) and at the end of
Sec.~\ref{sec:nonrelEFT}. The higher-order time-derivative terms
do not introduce new degrees of freedom, because such terms must
be considered perturbations around the zeroth-order equations. In
fact, one can use the equations of motion to remove the
higher-order time-derivative terms and replace them with
lower-order terms \cite{GrosseKnetter94}. (See also
\cite{Georgi91,Arzt95}.) We demonstrate this explicitly in the
following subsection for $\chi$ defined via the local field
redefinition of Eq.~(\ref{phipsi_loc}), in the free-field limit
(taking $\lambda \rightarrow 0$). This limit is sufficiently
nontrivial that a generalization to the interacting case would
not be straightforward. We also show that the final result, after
removing the higher-order time derivatives, is of exactly the
form one would obtain for $\psi$, as defined via the nonlocal
field redefinition. We pursue the comparison up to order
$\epsilon_x^5 \sim (k/m)^{10}$. 

\subsection{Higher-order time-derivative terms in the free-field limit }
\label{sec:local_free}

In the free-field limit the equation of motion for $\chi$,
Eq.~(\ref{eomG_loc}), simplifies to
\ba
\label{psieom_loc_free}
i \dot \chi = -\dfrac{1}{2m} \nabla^2 \chi - \frac{ e^{2imt}
}{2m} \nabla^2 \chi^*.
\ea 
Eq.~(\ref{psieom_loc_free}) (together with its complex conjugate)
can be solved exactly, with a solution of the form of
Eq.~(\ref{psichidefa}).  As discussed above, the solution is the
sum of two terms, one that varies slowly, and the other that
varies rapidly.  Thus it is straightforward to write an exact
expression for the slow mode, Eq.~(\ref{chispsis}).  However,
here we will put aside the exact solution, and use this case to
further illustrate the iterative approximation technique, as was
used in Appendix \ref{sec:local_EFT}. 

After decomposing the field $\chi$ into modes as in
Eq.~(\ref{psinu1}) and again assigning $\chi_{\nu = 0} \equiv
\chi_s$, the equation of motion for $\chi_s$ takes the form
\beqn
i \dot{\chi}_s = -\dfrac{1}{2m} \nabla^2 \chi_s -\dfrac{1}{2m}
\nabla^2\chi_{2}^*.
\label{eompsislow_loc_free}
\eeqn
Therefore, among all modes $\chi_\nu$ with $\nu \neq 0$, only
$\chi_2$ contributes to the evolution of $\chi_s$ in the limit
$\lambda \rightarrow 0$. So we only need to find an appropriate
substitution for $\chi_2$ in order to obtain the EFT for $\chi_s$
in this limit. From Eq.~(\ref{eompsinu_locn}) we may write
\beq
i \dot{\chi}_2 - 2 m \chi_2 = -\dfrac{1}{2m} \nabla^2 \chi_2
-\dfrac{1}{2m} \nabla^2 \chi_{s}^*.
\label{eompsi2_loc}
\eeq
Decomposing $\chi_2$ as in Eq.~\eqref{psinuepsilon_loc} it is
easy to obtain
\ba 
\chi_2^{(1)}=\dfrac{1}{4m^2} \nabla^2 \chi_s^* ,
\ea  
while for higher-order iterations we have
\ba 
\chi_2^{(n)}=\dfrac{1}{2m}\left( \dfrac{1}{2m} \nabla^2 +i \partial_t \right) \chi_2^{(n-1)}
\, , \quad {\text{for}} \quad n\geq 2
\ea 
all of which can be encapsulated in the relation
\ba 
\chi_2^{(n)}=\dfrac{1}{4m^2} \nabla^2 \left( \dfrac{1}{4m^2} \nabla^2 + \dfrac{i}{2m} \partial_t \right)^{n-1} \chi_s^*.
\ea 
Having obtained $\chi_2$ at arbitrary iteration $n$, one may try
to formally resum the infinite series to obtain
\ba 
\chi_2 =\sum_{n=1}^\infty \chi_2^{(n)}= \dfrac{1}{4m^2} \nabla^2  \left(1- \dfrac{1}{4m^2} \nabla^2 -  \dfrac{i}{2m} \partial_t \right)^{-1} \chi_s^*.
\ea 
Substituting this relation back to
Eq.~\eqref{eompsislow_loc_free} yields
\ba 
\label{eompsislow_full}
i \dot{\chi}_s = -\dfrac{1}{2m} \nabla^2 \chi_s -\dfrac{1}{8m^3}
\nabla^4 \left(1- \dfrac{1}{4m^2} \nabla^2+ \dfrac{i}{2m}
\partial_t \right)^{-1} \chi_s.
\ea 
Although the infinite series has been resummed, one can check
that the exact solution of Eq.~\eqref{eompsislow_loc_free} would
satisfy Eq.~\eqref{eompsislow_full}. Here we will only consider
the first few terms among the infinite tower of terms in order to
demonstrate how higher-order time derivatives enter the
calculation and may be systematically removed. To do so, it is
easiest to work in Fourier space, within which we expand
Eq.~(\ref{eompsislow_full}) up to the order $(k/m)^{10}$:
\ba 
\label{psi_s_free_k_10}
\nonumber i\dot \chi_s &\simeq& \dfrac{k^2}{2m} \chi_s 
\\ \nonumber 
&-&\dfrac{k^4}{8m^3}\chi_s
\\  
&+& \dfrac{ik^4}{16m^4}\dot \chi_s +\dfrac{k^6}{32m^5}\chi_s
\\ \nonumber 
&+&\dfrac{k^4}{32m^5} \ddot{\chi}_s -\dfrac{ik^6}{32m^6}\dot
\chi_s -\dfrac{k^8}{2^7m^7}\chi_s
\\ \nonumber 
&-&\dfrac{ik^4}{64m^6}\dddot{\chi}_s
-\dfrac{3k^6}{2^7m^7}\ddot{\chi}_s +\dfrac{3ik^8}{2^8m^8}\dot
\chi_s +\dfrac{k^{10}}{2^9m^9}\chi_s
\\ \nonumber 
&+&\calO[(k/m)^{12}] .
\ea 
In Eq.~(\ref{psi_s_free_k_10}) we have arranged terms of similar
magnitude to be in the same line. Notice that from the equation
of motion, Eq.~(\ref{eompsislow_loc_free}), one can conclude that
the time-derivative operator and the spatial Laplacian are at the
same order (i.e., in Fourier space, $\partial_t \sim k^2/m$) so
that, effectively, our expansion is in powers of $(k/m)$. 

Eq.~(\ref{psi_s_free_k_10}) is a higher-order equation in time
derivatives. But, as mentioned above, since all terms from the
second to the fifth line should be considered as perturbations,
we can replace higher-order time-derivative terms with
lower-order terms, upon using the equation of motion and its time
derivatives. The procedure is simple. First, consider
Eq.~\eqref{psi_s_free_k_10} as an algebraic equation for $\dot
\chi_s$; move all terms involving $\dot \chi_s$ to the left-hand
side; divide both sides by the resulting coefficient:
\beq
\left( 1-\dfrac{k^4}{16m^4}+\dfrac{k^6}{32m^6}-\dfrac{3k^8}{2^8m^8} \right) ,
\eeq
and expand up to the desired order (here $\calO(k/m)^{10}$). This
yields
\ba 
\label{psi_s_red_1}
\nonumber i\dot \chi_s &\simeq& \dfrac{k^2}{2m} \chi_s 
\\ \nonumber 
&-&\dfrac{k^4}{8m^3}\chi_s
\\ 
&+&\dfrac{k^6}{16m^5}\chi_s
\\ \nonumber 
&+&\dfrac{k^4}{32m^5} \ddot{\chi}_s -\dfrac{k^8}{32m^7}\chi_s
\\ \nonumber 
&-&\dfrac{ik^4}{64m^6}\dddot{\chi}_s
-\dfrac{3k^6}{2^7m^7}\ddot{\chi}_s +\dfrac{k^{10}}{64m^9}\chi_s
\\
\nonumber &+& {\cal O} [ (k / m)^{12} ] .
\ea 
The right-hand side of Eq.~(\ref{psi_s_red_1}) contains no first
time-derivative terms, though it still contains higher-order
ones. To remove the $\ddot \chi_s$ terms, we take a time
derivative of Eq.~\eqref{psi_s_red_1} once and remove terms
containing $\dot \chi_s$ using Eq.~\eqref{psi_s_red_1} itself,
again only keeping terms up to the desired order in $(k / m)$.
This yields an algebraic equation for $\ddot \chi_s$ that does
not contain $\dot \chi_s$. Using the resulting equation to remove
$\ddot \chi_s$ from Eq.~\eqref{psi_s_red_1} yields
\ba 
\label{psi_s_red_2}
\nonumber i\dot \chi_s &\simeq& \dfrac{k^2}{2m} \chi_s 
\\ \nonumber 
&-&\dfrac{k^4}{8m^3}\chi_s
\\ 
&+&\dfrac{k^6}{16m^5}\chi_s
\\ \nonumber 
&-& \dfrac{5k^8}{2^7m^7}\chi_s
\\ \nonumber 
&-&\dfrac{ik^4}{64m^6}\dddot{\chi}_s
+\dfrac{13k^{10}}{2^9m^9}\chi_s.
\ea 
By this series of steps, we have removed $\ddot \chi_s$ from the
equation. We repeat the same procedure to remove $\dddot \chi_s$
by taking the time derivative of Eq.~(\ref{psi_s_red_2}) twice
and substituting the resulting expression for $\dddot \chi_s$
into Eq.~\eqref{psi_s_red_2}. This finally yields
\ba 
\label{psi_s_red_f}
i\dot \chi_s = \left[ \dfrac{k^2}{2m}
-\dfrac{k^4}{8m^3}+\dfrac{k^6}{16m^4}-\dfrac{5k^8}{128m^7}+\dfrac{7k^{10}}{256m^9}
\right]\chi_s + \calO [ ( k / m )^{12} ] ,
\ea 
which no longer includes any higher-order time-derivative terms.
The form of Eq.~(\ref{psi_s_red_f}) is in complete agreement (up
to the working order) with the expansion of the equation of
motion in the free-field limit that one would obtain by using the
nonlocal field redefinition, namely
\beq
i \dot{\psi}_s = m \left( \sqrt{ 1 + \frac{ k^2}{m^2} } - 1
\right) \psi_s ,
\eeq
which is the Fourier transform of Eq.~\eqref{eompsis1} in the
limit $\lambda \rightarrow 0$. 

\section{Comparison with the Recent Calculation by Mukaida, Takimoto, and Yamada  } 
\label{sec:Masaki_compare}

In this appendix we compare our EFT with the recent calculation
by Mukaida, Takimoto, and Yamada in Ref.~\cite{MasakiOscillons}.
We show that the two resulting low-energy descriptions are
equivalent, related by a field redefinition. 

In Ref.~\cite{MasakiOscillons} the authors begin by separating
the real-valued scalar field of the relativistic theory, $\phi$,
into a nonrelativistic component and fluctuations, $\phi (x) =
\phi_{\rm NR} (x) + \delta \phi (x)$. The nonrelativistic field
$\phi_{\rm NR}$ is defined as
\beq
\phi_{\rm NR} (x) = \int_{K \in {\rm NR}} d^4 K e^{-i K \cdot x} \phi (K) ,
\label{phiNRdef}
\eeq
where $K \in {\rm NR}$ indicates the region of four-momentum
$K^\mu = (k^0, {\bf k})$ with $\vert k^0 \vert \sim m c^2 + {\cal
O} (m v^2)$ and $0 \leq \vert {\bf k} \vert \leq m v$, with $v
\ll c$. (The fluctuation $\delta \phi$ is defined as the Fourier
integral over the complementary range of $K^\mu$.) Next the
authors parameterize the nonrelativistic component as
\beq
\phi_{\rm NR} (t, {\bf x}) = \frac{1}{2} \left[ \sigma (t, {\bf x} ) e^{-imt} + \sigma^* (t, {\bf x}) e^{i mt} \right] .
\label{phiNRsigma}
\eeq
(The field we have labeled $\sigma$ is denoted by $\Psi$ in
Ref.~\cite{MasakiOscillons}; we use $\sigma$ to avoid confusion
with the modes $\Psi_\nu$ defined in Eq.~(\ref{Psinu3}).) In
contrast with our approach, the authors of
Ref.~\cite{MasakiOscillons} first separate the relativistic field
$\phi$ into components with small and large spatial momenta,
$\sigma$ and $\delta \phi$ (respectively), and then construct an
EFT for $\sigma$, whereas we relate the real-valued relativistic
field $\phi$ to the complex field $\psi$ via Eqs.~(\ref{phipsi1})
and construct an EFT for the slowly varying portion, $\psi_s$,
identified as the $\nu = 0$ mode of Eq.~(\ref{psinu}).  This
suggests that the two resulting low-energy effective descriptions
might be related by a field redefinition.  In particular,
substituting the mode decomposition of Eq.~\eqref{psinu} into
Eq.~\eqref{phipsi1}, we may relate $\sigma$ to $\psi$:
\beq
 \frac{1}{2}  \left[  \sigma (t, {\bf x} ) e^{-imt} + \sigma^* (t, {\bf x}) e^{i mt} \right] = \frac{1}{ \sqrt{ 2m} } {\cal P}^{-1/2} \left[ \left( \psi_s + \psi_2^* \right) e^{-imt} + \left( \psi_s^* + \psi_2 \right) e^{imt} \right] .
\label{phiNRpsi}
\eeq
Since $\sigma$, $\psi_s$, and $\psi_2$ are each slowly varying
functions of $t$, Eq.~(\ref{phiNRpsi}) implies that
\ba 
\label{redef_full}
\sqrt{\dfrac m2} \, \sigma =\calP^{-1/2}( \psi_s + \psi_2^*)
\ .
\ea 
At leading order in $\epsilon_x, \epsilon_t$, and $\lambda$, this
reduces to
\ba 
\sqrt{\dfrac m2}\, \sigma = \left(1+\dfrac{1}{4m^2} \nabla^2 \right) \psi_s -\dfrac{\lambda}{16m^3} \vert \psi_s \vert^2 \psi_s  + {\cal O} (\epsilon^2) ,
\label{redef}
\ea 
where we have expanded the nonlocal operator ${\cal P}$, used the
relations in Eqs.~\eqref{Psinu1} and \eqref{G0G1} to replace
$\psi_2^*$, and adopted the notation ${\cal O} (\epsilon^2)$ of
Eq.~(\ref{Orderepsilon}) to indicate terms that are at least
second order in the small quantities $\epsilon_x$, $\epsilon_t$,
and $\lambda$. 

Before showing the equivalence of the low-energy effective
descriptions for $\sigma$ and $\psi_s$, we note that one can
derive the same relation as in Eq.~\eqref{redef}, starting from
the locally defined field $\chi$ of Eq.~\eqref{phipsi_loc}. In
that case, the relation between the two fields becomes
$\sqrt{m/2} \, \sigma = \chi_s + \chi_2^*$, which differs from
Eq.~\eqref{redef_full} by the absence of the nonlocal operator
${\cal P}$. However, according to Eq.~\eqref{psinu1_loc}, the
$\chi_2^*$ term acquires an extra contribution that exactly
compensates the missing term due to the absence of ${\cal P}$
from the relation between $\sigma$ and $\chi$. 

To demonstrate that the low-energy effective descriptions for
$\sigma$ and $\psi_s$ are indeed related through the field
redefinition of Eq.~\eqref{redef}, we first consider the
equations of motion. In Eq.~(2.15) of Ref.~\cite{MasakiOscillons}
the authors present their effective Lagrangian, which is given by
\ba 
\label{L_Masaki}
{\cal L} = \dfrac{1}{4} \left[2im \sigma^* \dot \sigma -\sigma^*
\ddot \sigma + \sigma^* \nabla^2 \sigma -V_{\text{eff}}\right] ,
\ea 
where, from Eq.~(2.17) of Ref.~\cite{MasakiOscillons},
\ba 
V_{\text{eff}} (\sigma, \sigma^*) = -\dfrac{3 g_4}{8} \vert
\sigma \vert^4 + \dfrac{g_4^2}{128m^2} \vert \sigma \vert^6 .
\ea 
Here $g_4$ is the coupling constant of the quartic interaction in
the original, relativistic theory, and we have set the cubic
interaction to zero, to match the form of $V (\phi)$ that we have
considered throughout our analysis. The quartic coupling constant
$g_4$ of Ref.~\cite{MasakiOscillons} is related to the coupling
$\lambda$ we introduced in Eq.~(\ref{Lrel}) by $g_4= -\lambda/6$.

The equation of motion for $\sigma$, up to the order that matches
our analysis, can be obtained by varying the action with respect
to $\sigma^*$, which yields
\ba 
i\dot \sigma = \dfrac{1}{2m} \ddot \sigma -\dfrac{1}{2m}\nabla^2
\sigma +\dfrac{1}{2m}V_{\text{eff},\sigma^*} .
\label{eomsigma}
\ea 
Eq.~(\ref{eomsigma}) is second order in time derivatives. Upon
substituting the field redefinition of Eq.~(\ref{redef}), we find
an equation of motion for $\psi_s$ which is also second order in
time derivatives:
 \beqn
  \begin{split}
i\dot{\psi}_s &= - \frac{1}{2m} \nabla^2 \psi_s + \frac{
\lambda}{8 m^2} \vert \psi_s \vert^2 \psi_s \\ &\quad
+\dfrac{1}{2m}\ddot \psi_s-\dfrac{i}{4m^2}\nabla^2 \dot \psi_s
+\dfrac{i\lambda}{16m^3} \psi_s\left( 2 \dot \psi_s
\psi_s^*+\psi_s \dot \psi_s^* \right)
  \\ 
&\quad -\frac{1}{8 m^3} \nabla^4 \psi_s+ \frac{ \lambda}{32 m^4}
\bigg[ \psi_s^2 \nabla^2 \psi_s^* + 2 \vert \psi_s \vert^2
\nabla^2 \psi_s + \nabla^2 \left( \vert \psi_s \vert^2 \psi_s
\right) \bigg] - \frac{17 \lambda^2}{768 m^5} \vert \psi_s
\vert^4 \psi_s + {\cal O} (\epsilon^3 ).
  \end{split}
  \label{psiseom_prototype}
  \eeqn
Eq.~(\ref{psiseom_prototype}) does not appear to match the
equation of motion for $\psi_s$ which we found above, in
Eq.~(\ref{eompsislow}). In particular, the two equations differ
by the presence of the terms in the second line of
Eq.~(\ref{psiseom_prototype}). However, to the perturbative order
to which we are working, it is straightforward to show that the
new terms that appear in Eq.~(\ref{psiseom_prototype}) vanish. In
particular, we may take one time derivative of
Eq.~(\ref{psiseom_prototype}) to find
\beq
i \ddot{\psi}_s = - \frac{1}{ 2m} \nabla^2 \dot{\psi}_s + \frac{
\lambda}{8 m^2} \psi_s \left( 2 \dot{\psi}_s \psi_s^* + \psi_s
\dot{\psi}_s^* \right) + {\cal O} (\epsilon^3 ) .
\label{psieomnewterms}
\eeq
Substituting Eq.~(\ref{psieomnewterms}) into
Eq.~(\ref{psiseom_prototype}) yields
 \beqn
 \begin{split}
i\dot{\psi}_s &= - \frac{1}{2m} \nabla^2 \psi_s + \frac{
\lambda}{8 m^2} \vert \psi_s \vert^2 \psi_s \\ &\quad\quad -
\frac{1}{8 m^3} \nabla^4 \psi_s + \frac{ \lambda}{32 m^4} \bigg[
\psi_s^2 \nabla^2 \psi_s^* + 2 \vert \psi_s \vert^2 \nabla^2
\psi_s + \nabla^2 \left( \vert \psi_s \vert^2 \psi_s \right)
\bigg] - \frac{17 \lambda^2}{768 m^5} \vert \psi_s \vert^4 \psi_s
+ {\cal O} (\epsilon^3 ) .
 \end{split}
 \label{psiseomC}
 \eeqn
To the order to which we have been working, Eq.~(\ref{psiseomC})
exactly matches Eq.~(\ref{eompsislow}). We therefore find that
the equations of motion for the two low-energy effective
descriptions are indeed equivalent, at least up to the working
order. 

So far our discussion has established the equivalence between the
two low-energy descriptions at the classical level. To analyze
the equivalence even for matters concerning quantization, we next
consider the relevant Lagrangians. Substituting the field
redefinition of Eq.~(\ref{redef}) into the Lagrangian of
Eq.~(\ref{L_Masaki}) and performing some straightforward algebra,
we find
\ba 
\begin{split}
{\cal L}[\psi_s] &= \frac{i}{2} \left( \dot{\psi}_s \psi_s^* -
\psi_s \dot{\psi}_s^* \right) - \frac{1}{2m} \nabla \psi_s \nabla
\psi_s^* - \frac{ \lambda}{16 m^2} \vert \psi_s \vert^4 \\ &\quad
-\dfrac{1 }{2m} \psi_s^* \, \ddot \psi_s - \dfrac{i }{2m^2} \dot
\psi_s^* \, \nabla^2 \psi_s+ \frac{1 }{ 4 m^3} \nabla^2 \psi_s
\nabla^2 \psi_s^* - \dfrac{i\lambda}{8m^3} \vert \psi_s \vert^2
\psi_s^* \dot \psi_s
\\
&\quad - \frac{ \lambda}{16 m^4} \vert \psi_s \vert^2 \left(
\psi_s^* \nabla^2 \psi_s + \psi_s \nabla^2 \psi_s^* \right) +
\frac{35 \lambda^2}{9 \times 2^8 \, m^5} \vert \psi_s \vert^6,
\end{split}
\label{Lpsislow_2}
\ea 
where we have neglected some boundary terms. The Lagrangian of
Eq.~(\ref{Lpsislow_2}) is different from the one we obtained in
Eq.~\eqref{Lpsislow}, although it is easy to show that
Eq.~(\ref{Lpsislow_2}) gives rise to the same equations of
motion. To demonstrate the equivalence between the Lagrangians in
Eqs.~(\ref{Lpsislow}) and (\ref{Lpsislow_2}), we perform another
field redefinition:
\ba 
\psi_s =\tilde \psi    -\dfrac{1}{4m} \left( i \dot {\tilde \psi} + \frac{1}{2m} \nabla^2 \tilde \psi - \frac{ \lambda}{8 m^2} \vert \tilde \psi \vert^2 \tilde \psi \right) .
\label{redef_2}
\ea  
Substituting Eq.~(\ref{redef_2}) into Eq.~(\ref{Lpsislow_2})
yields
\beqn 
\begin{split}
{\cal L}[\tilde \psi ]&= \frac{i}{2} \left( \dot{\tilde \psi}
\tilde \psi^* - \tilde \psi \dot{\tilde \psi}^* \right) -
\frac{1}{2m} \nabla \tilde \psi \nabla \tilde \psi^* - \frac{
\lambda}{16 m^2} \vert \tilde \psi \vert^4 \\ &\quad\quad +
\frac{1}{ 8 m^3} \nabla^2 \tilde \psi \nabla^2 \tilde \psi^* -
\frac{ \lambda}{32 m^4} \vert \tilde \psi \vert^2 \left( \tilde
\psi^* \nabla^2 \tilde \psi + \tilde \psi \nabla^2 \tilde \psi^*
\right) + \frac{17 \lambda^2}{9 \times 2^8 \, m^5} \vert \tilde
\psi \vert^6.
\end{split}
\label{Lpsislow_3}
\eeqn
This is precisely the form of the Lagrangian of
Eq.~\eqref{Lpsislow}, after trivial relabeling of the dynamical
field $\psi_s \to \tilde \psi$. Since the Lagrangians of
Eqs.~(\ref{L_Masaki}), (\ref{Lpsislow_2}), and (\ref{Lpsislow_3})
are related to each other via field redefinitions, we expect
their corresponding $S$-matrices to remain equivalent as well, at
least up to the perturbative order to which we have been working
\cite{Georgi91,GrosseKnetter94,Arzt95}.

We close with some comments on the field redefinitions in
Eqs.~(\ref{redef}) and (\ref{redef_2}). Both of the field
redefinitions may be written in the form $\psi \rightarrow \psi +
\epsilon T (\psi)$, where $T (\psi)$ is a local function of
$\psi$ and its derivatives. As demonstrated by Arzt in
Ref.~\cite{Arzt95} (see also
Refs.~\cite{Georgi91,GrosseKnetter94}), within an EFT context,
such redefinitions do not change the $S$-matrix. We therefore
conclude that the low-energy effective description we have
derived for the field $\psi_s$ is equivalent, to our working
order in perturbation theory, to the low-energy description
derived in Ref.~\cite{MasakiOscillons}, at both the classical and
quantum levels \cite{FRBnote}.

\section*{Acknowledgements}

It is a pleasure to thank Mustafa Amin, Mark Hertzberg, Chanda
Prescod-Weinstein, Iain Stewart, and Masaki Yamada for helpful
discussions, and to acknowledge support from the Center for
Theoretical Physics at MIT. This work is supported in part by the
U.S. Department of Energy under grant Contract Number
DE-SC0012567.


\begin{thebibliography}{9999}
\bibitem{Virialfn} Axion dark matter is expected to be virialized in galaxies, though in that case the dominant contribution to the potential energy arises from gravitational self-interactions rather than from the $\lambda \phi^4$ interaction that we study here. See, e.g., Refs.~\cite{Zwicky33,BinneyTremaine}.
\end{thebibliography}
\end{document}